\documentstyle[prl,aps]{revtex}
\input{psfig}
\title{Long-range correlations in non-equilibrium systems:\\ 
 Lattice gas automaton approach.}
\author{Alberto Su\'arez and Jean Pierre Boon \\
Center for Nonlinear Phenomena and Complex Systems\\ 
Universit\'e Libre de Bruxelles - Campus Plaine - C.P. 231 \\ 
1050 Brussels, Belgium \\
Patrick Grosfils \\
Department of Applied Physics, Tokyo Institute of Technology \\ 
Tokyo 152, Japan}
\date{To appear in Physical Review E (1996).}
\begin{document}
\maketitle
\vspace*{1cm}
\begin{abstract}
In systems removed  from equilibrium, intrinsic microscopic fluctuations become correlated over distances comparable to the characteristic macroscopic length over which the external  constraint is exerted. In order to investigate this phenomenon,  we construct a microscopic model with simple stochastic dynamics using lattice gas automaton rules that satisfy local detailed balance. Because of the simplicity of the automaton dynamics,   analytical theory can be developed to describe the space and time evolution of the density fluctuations. The exact equations for the pair correlations are solved explicitly in the hydrodynamic limit. In this limit, we rigorously derive  the results obtained phenomenologically by fluctuating hydrodynamics. In particular, the spatial algebraic decay of the equal-time fluctuation correlations predicted by this theory  is found to be in excellent agreement with the results of our lattice gas automaton simulations for two different types of boundary  conditions. Long-range correlations of the type described here appear generically in dynamical systems that exhibit large scale anisotropy and lack detailed balance. 
\end{abstract}
\par
\begin{center}
\end{center}
\section{Introduction}
In a hydrodynamic system under non-equilibrium conditions the fluctuations of the densities of conserved quantities  are correlated over  large distances, as confirmed by recent experiments performed by Law, Sengers {\it et al} \cite{law89,law90,segre92,li94}. The presence of long-range correlations in  systems removed from equilibrium had been predicted by kinetic theory \cite{kirkpatrick82}, by non-equilibrium statistical mechanics \cite{ronis80}, and by phenomenological theories, such as fluctuating hydrodynamics \cite{ronis82}. These  correlations decay algebraically over distances comparable to the size of the system. They appear generically in systems subject to non-equilibrium constraints \cite{ronis80,fox82,tremblay84,perezmadrid86,hwa89,garrido90,beijeren90,grinstein91,pagonabarraga94,dorfman94} and are a consequence of the existence of conserved quantities, the absence of detailed balance, and the presence of spatial anisotropy. The long-range nature of these correlations is  remarkable in as much 
as it is manifested in fluids where the interactions are short-ranged, and under conditions for which the fluid is far from critical points or hydrodynamic instabilities. The fact that equilibrium  correlations remain short-ranged away from critical points is a consequence of precise cancelations of the effects of noise sources \cite{ronis80}. As soon as this balance, which is characteristic of the equilibrium state, is lost the correlations may become long-ranged, as they generically do.\\
\indent Our objective in the present work is to put forth and  analyze a simple microscopic model, which nonetheless possesses sufficiently complex dynamics to exhibit this type of  long-range correlations. In particular we construct a lattice gas automaton (LGA) corresponding to a collection of random walkers. The ``particles'' move on a regular array at discrete time intervals and interact via an exclusion principle, a constraint which acts as a sort of hard-core potential in the lattice. Particles also enter collisions whose outcome, while conserving the number of particles, is otherwise entirely random. These collision rules satisfy a local  detailed balance relation, and  conserve momentum globally, in a statistical way, but not locally. At a global scale, detailed balance is absent because of the imposition of non-equilibrium constraints. The automaton dynamics naturally lends itself to a hierarchical  description\cite{lebowitz88,grosfils92,bussemaker95}. Microscopically, particles propagate between adjacent nodes and experience collisions. From a ``macroscopic'' point of view, the  evolution of the automaton is given by a diffusion equation. An intermediate ``mesoscopic'' description accounts for the statistical properties of the fluctuations, which correspond to deviations from the average behavior arising from the  microdynamics. In equilibrium, the fluctuation correlations are localized on  a single node.  In the presence of a density gradient, which maintains the automaton away from equilibrium,  we can distinguish two contributions in the correlation function of the  particle number fluctuations: A local equilibrium one, which is short-ranged (in our model, it is strictly localized  on one lattice node), and a long-range term, which decays algebraically with a characteristic length on the order of the system size. The simplicity of the dynamics of this automaton at the microscopic level makes it possible to develop an analytic description not only for the evolution of the densities of conserved quantities of the automaton (in this case, solely the number of particles per node), but also for their fluctuations. Thus, we derive exact equations for the evolution of pair correlations and solve them in the hydrodynamic limit.   The lowest order approximation in a  perturbative scheme using the inverse of the size of the system as an expansion parameter, leads to the same expressions for  pair correlations as the phenomenological theory of fluctuating hydrodynamics. The validity of this  approximation is  tested against the results of simulations of the dynamics of the automaton. It is seen that even for very small automata (that is, even rather far from the thermodynamic limit)  fluctuating hydrodynamics provides a very accurate description of the statistical properties of the fluctuations.\\ 
\indent The present model system  is similar to a cellular automaton  proposed by Kawasaki \cite{kawasaki66} and studied by Spohn \cite{spohn83}, who also derived the existence long-range correlations from the  automaton microscopic dynamics. From a mesoscopic viewpoint, both the cellular automaton studied by Spohn and the lattice gas automaton (LGA) investigated here represent two different microscopic realizations of a stochastic equation analogous to that studied by Nicolis and Malek-Mansour \cite{nicolis84} and Garcia {\it et al} \cite{garcia87} to describe heat transport in a rigid conductor subject to a temperature gradient. In this case, the conserved quantity is the energy density (as measured by the local temperature), and the temperature fluctuations are described by a Fourier (heat diffusion) equation, to which a random heat current with a local equilibrium form is added. \\
\indent Long-range correlations  of the kind described in this paper are also present in lattice gas automata (LGA) with collision rules that violate detailed balance. These automata attain a homogeneous equilibrium state which is non-Gibbsian. Initially automata violating detailed balance were constructed to simulate hydrodynamics at  high Reynolds numbers  \cite{dubrulle90}. It was  realized later on that these models are intrinsically interesting as a paradigm for driven systems \cite{bussemaker95,ernst95}, and that they exhibit all the wealth of behavior characteristic of systems removed from equilibrium. In particular, they also exhibit algebraically decaying correlations, which have been studied in great detail \cite{garrido90,bussemaker95,ernst95}.\\
\indent Our work complements that carried out by these authors and  provides a systematic comparison between the theoretical description and  simulations. One of the main objectives of this paper is to derive the exact equations for the evolution of the hydrodynamic variables and their fluctuations and to establish the connection with  fluctuating hydrodynamics. Once this program is realized, we can  describe in detail the mechanism by which the long-range correlations are built up from the local microdynamics. We also discuss the validity of the local-equilibrium hypothesis, a basic assumption (usually justified {\it a posteriori}) in the theory of fluctuating hydrodynamics.\\
\indent In section II, we review   the problem of heat transport in a rigid conductor using the phenomenological approach of fluctuating hydrodynamics. The novelty of the present treatment is that the postulated Landau equation includes  explicitly  the heat reservoirs that maintain the temperature gradient across the conductor. In this scheme  it is possible to discuss the effect of boundary conditions in the stochastic equation  rigorously. Given the long-range nature of the correlations,  boundary effects should be non-trivial. It is  argued that the implemented boundary conditions for the automaton described in section III (vanishing of long-range correlations at the boundaries of the system) correspond to the paradigm of a system in a quasi-stationary steady state maintained by contact (diffusive or thermal) with reservoirs. Section III constitutes the main body of the paper. We construct a two-dimensional lattice gas automaton whose collision rules satisfy a local detailed balance relation. Global detailed balance is broken by imposing non-equilibrium constraints. The equations for the evolution of the average number of particles per node and for the corresponding fluctuations are derived from the microscopic propagation and collision rules. Simulations of the dynamics of the automaton demonstrate the accuracy of the theoretical description. Section IV contains a summary of results and some concluding comments. 
\section{Fluctuating hydrodynamics.}
We consider the problem of heat transport in a one-dimensional rigid conductor. Our  starting point is the isolated system analyzed by Procaccia {\it et al} in Appendix B of Ref.\cite{procaccia79} (see Fig. 1). We  proceed in successive  steps: First,  the heat diffusion equation for the system depicted in Fig. 1 is solved. Then, we discuss the conditions under which such system supports a quasi-stationary non-equilibrium steady state (quasi-NESS) with a linear temperature profile for the central portion. Finally, we solve the stochastic equation, which is constructed by adding a random heat flux with  local equilibrium form  to the Fourier equation, in the limit in which the quasi-NESS is established.\\
\begin{figure}[h]
\[\psfig{figure=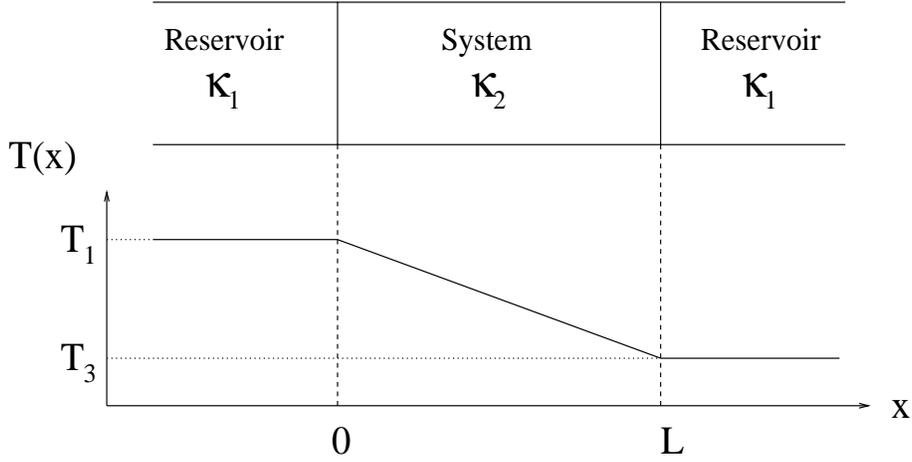,width=12cm,height=6cm}\]
\caption{Quasi-NESS temperature profile.} 
\end{figure} 
\indent Consider the Fourier equation in one dimension,
\begin{equation}
\frac{\partial ~}{\partial t} ~ T(x,t) ~ = ~ \frac{\partial ~}{\partial x} \left[ \kappa(x) \frac{\partial ~}{\partial x} ~ T(x,t) \right],
\end{equation}
where $~ \kappa(x) ~ = ~   \kappa_1 ~ (\theta(-x) + \theta(x-L))  ~+~ \kappa_2 ~\theta(x) ~ \theta(L-x) ~$ is the piecewise constant thermal diffusivity ($~ \theta(x)~$ is the Heaviside step function). The initial condition is 
\begin{equation}
T(x,0) ~ = ~  T_1 ~\theta(-x)  ~ +~ T_2 ~\theta(x) ~ \theta(L-x) ~ + T_3 ~ \theta(x-L),
\end{equation}
where $~ T_2 ~ = ~ \frac{T_1+T_3}{2}~ = ~ T_1 + \beta \frac{L}{2} ~,$ and $~ \beta  =  \frac{T_3-T_1}{L}~$ is the magnitude of the temperature gradient, and a measure of how far from equilibrium the system is.\\
\indent Rather than directly solving Eq. (1), we consider the equivalent equation for the spatial derivative of the temperature profile
\begin{equation}
\frac{\partial ~}{\partial t} \frac{\partial T(x,t)}{\partial x} ~ = ~ \frac{\partial^2 ~}{\partial x^2} ~ \left[ \kappa(x)  \frac{\partial T(x,t)}{\partial x} \right],
\end{equation}
with the corresponding initial condition
\begin{equation}
\frac{\partial T(x,0)}{\partial x} ~ = ~ \beta ~\frac{L}{2} ~ \left[ \delta(x) ~ + ~ \delta(x-L) \right].
\end{equation}
The formal solution of Eq. (3)  is given by
\begin{equation}
\frac{\partial  T(x,t)}{\partial x} ~ = ~ \exp \left\{  \frac{\partial^2 ~}{\partial x^2} ~ \kappa(x) ~t \right\} ~ \frac{\partial  T(x,0)}{\partial x}.
\end{equation} 
The following step is to find the eigenfunctions $~P_k(x)~$ of the operator $~ \frac{\partial^2 ~}{\partial x^2}\kappa(x) ~$:
\begin{equation}
\frac{\partial^2 ~}{\partial x^2} ~\kappa(x) ~ P_k(x) ~ = ~ -~k^2 ~ P_k(x), 
\end{equation}
which are given explicitly in Appendix A. Here, we just quote the result for  $~P_k(x)~$ in the limit $~ \epsilon ~= ~ \sqrt{\frac{\kappa_2}{\kappa_1}} ~\rightarrow~ 0 ~$ for the case of semi-infinite reservoirs:
\begin{equation}
P_k(x) ~ \approx ~ \frac{a_k}{\kappa_2} ~ \cos{\frac{kx}{\sqrt{\kappa_2}}} ~ \theta(x) ~ \theta(L-x) ~ + ~ {\cal{O}}(\epsilon), 
\end{equation}
and
\begin{equation}
a_k^2 ~ {\sim} ~ \frac{\kappa_2}{L} ~ \sum_{n=-\infty}^{\infty} ~ \delta(k ~-~ n \frac{\pi \sqrt{\kappa_2}}{L}), 
\end{equation}
when $~ \epsilon \rightarrow 0.$\\
\indent Using the spectral decomposition of the identity
\begin{equation}
\delta(x-x_0) ~ = ~ \int dk ~ \kappa(x_0) ~P_k(x_0)~ P_k(x),
\end{equation}
we can  give an explicit solution for Eq. (3) with the initial conditions given by Eq. (4)
\begin{eqnarray}
\frac{\partial  T(x,t)}{\partial x} & = & \beta ~ \frac{L}{2} ~ \exp\left\{ \frac{\partial^2 ~}{\partial x^2} \kappa(x) t \right\} ~ \left[ \delta(x-L) + \delta(x) \right] ~  \nonumber \\
& \approx & \frac{\beta}{2} ~ \left[~ \sum_{n=-\infty}^{\infty} ~ e^{-n^2 \frac{\pi^2 \kappa_2}{L^2} t } ~ \cos{\frac{n \pi x}{L}} ~ \left(1 + (-1)^n \right) ~ \right] ~ \theta(x) ~ \theta(L-x).
\end{eqnarray} 
This solution is  valid for $ ~ t \ll \frac{1}{\epsilon^2} ~ \frac{L^2}{\pi^2 \kappa_2} ~$ and $~ \epsilon  ~ \rightarrow ~ 0 $.\\
\indent We note that for an intermediate time regime
\[  \frac{L^2}{\pi^2 \kappa_2} \ll ~t~ \ll~ \frac{1}{\epsilon^2} ~ \frac{L^2}{\pi^2 \kappa_2}, \]
the derivative of the temperature profile is well approximated by
\begin{equation}
\frac{\partial ~}{\partial x}  ~T(x,t) ~ \approx ~ \beta ~ \theta(x) ~ \theta(L-x),
\end{equation}
which corresponds to  the long-lived quasi-NESS depicted in Fig. 1, previously derived in Ref. \cite{procaccia79}.\\
\indent In order to understand what are the proper boundary conditions in a non-equilibrium system when fluctuations are considered  we  now proceed to  solve the stochastic equation
\begin{equation}
\frac{\partial ~}{\partial t} ~ T(x,t) ~ = ~ \frac{\partial ~}{\partial x} \left[ \kappa(x) \frac{\partial ~}{\partial x}~ T(x,t) \right] ~ + ~ \frac{\partial ~}{\partial x}~ g(x,t),
\end{equation}
obtained by adding to the deterministic heat-diffusion equation (Eq. (1)) the gradient of a random heat flux $~ g(x,t)$, which is assumed to be Gaussian white noise with the properties:
\begin{eqnarray}
\left< g(x,t) \right> & = & 0,\nonumber \\
\left< g(x,t) g(x^{'},t^{'}) \right> & = & 2 ~\frac{k_B}{C_P} ~ \kappa(x) ~\left< T(x,t) \right>^2 ~ \delta(x-x^{'}) ~ \delta(t - t^{'}),
\end{eqnarray}
where $~ k_B ~$ is the Boltzmann constant, $~ C_p~$ the  heat capacity per unit volume, and $~ \left<T(x,t)~ \right>~$ is the temperature profile which is a solution of the deterministic equation, Eq. (1). This Landau equation (Eq. (12)) is constructed  phenomenologically  by requiring that at homogeneous equilibrium ($ \left<T(x,t)~ \right> =  T_{eq}$) the pair correlations are given correctly. The extension to a non-equilibrium situation makes use of the hypothesis of local equilibrium. It is assumed  that the thermodynamic variables that characterize the system are well defined locally. The non-equilibrium steady state is thus viewed as a state where these thermodynamic variables vary slowly in space (on a hydrodynamic scale, which is much larger than the scale at which the underlying microdynamics take place). In general, the derivation of Eq. (12), even in an approximate way, is rather complicated. In the following section we  analyze a diffusive lattice gas automaton for which we derive from the actual
 microscopic dynamics a  stochastic Landau equation.\\ 
\indent The equal time pair correlation function of the temperature fluctuations is defined as 
\begin{equation}
C(x,x^{'};t) ~= ~ \left< \delta T(x,t) ~ \delta T(x^{'},t) \right>,
\end{equation}
with $~ \delta T(x,t) ~ = ~ T(x,t) ~ - ~ \left<T(x,t) \right> $. This correlation function is the solution of the differential equation
\begin{eqnarray}
\frac{\partial ~}{\partial t} ~ C(x,x^{'};t)  & = & \left( \frac{\partial ~}{\partial x} \kappa(x) \frac{\partial ~}{\partial x} ~ + ~  \frac{\partial ~}{\partial x^{'}} \kappa(x^{'})  \frac{\partial ~}{\partial x^{'}}  \right) ~ C(x,x^{'};t)   ~ \nonumber \\
&   & \ \ \ \ \ \  \ \ \ \ \ +~ \frac{2 k_B}{C_p} ~ \frac{\partial ~}{\partial x} ~ \frac{\partial ~}{\partial x^{'}} ~ \kappa(x) ~ \left< T(x,t) \right>^2 ~ \delta(x-x^{'}). 
\end{eqnarray}
This equation is obtained by solving formally Eq. (12), 
constructing the equal time correlation function, and using
 (13) to evaluate one of the time integrals; 
differentiation of the resulting expression with respect to 
time  yields Eq. (15). \\
\indent There are two contributions to the correlations: a local-equilibrium contribution (noted $LE$), and the remainder (noted $LR$), which is long-ranged
\begin{equation}
C(x,x^{'};t) ~ = ~ C^{LE}(x,x^{'},t) ~ + ~ C^{LR}(x,x^{'},t),
\end{equation}
The local equilibrium contribution has the form
\begin{equation}
C^{LE}(x,x^{'};t) ~ = ~ \frac{k_B}{C_P} \left<T(x,t)\right>^2 ~ \delta(x-x^{'}),
\end{equation}
and the long-ranged term obeys the differential equation
\begin{eqnarray}
\frac{\partial ~}{\partial t} ~C^{LR}(x,x^{'};t)  & = & \left( \frac{\partial ~}{\partial x} \kappa(x) \frac{\partial ~}{\partial x} ~ + ~   \frac{\partial ~}{\partial x^{'}} \kappa(x^{'}) \frac{\partial ~}{\partial x^{'}}  \right) ~ C^{LR}(x,x^{'};t)  ~  \nonumber \\
& + &  \frac{2 k_B}{C_p} ~ \kappa(x) ~ \left( \frac{\partial \left< T(x,t) \right>}{\partial x} \right)^2  ~ \delta(x-x^{'}). 
\end{eqnarray}
Equation (18) corresponds to a diffusion equation in two dimensions with a source term on the line $~ x = x^{'} $, which is proportional to the square of spatial derivative of the temperature profile. Assuming that the system is initially in a local equilibrium state ($~C^{LR}(x,x^{'})  = 0 ~$), the formal solution of Eq. (18) is
\begin{eqnarray}
  C^{LR}(x,x^{'};t) & = & \frac{2 k_B}{C_p} ~ \int_0^t d\tau  ~\exp\left\{ \left(\frac{\partial ~}{\partial x} \kappa(x) \frac{\partial ~}{\partial x}  +    \frac{\partial ~}{\partial x^{'}} \kappa(x^{'}) \frac{\partial ~}{\partial x^{'}}  \right) (t-\tau) \right\} ~ \nonumber \\
&  \times & ~   \kappa(x) ~ \left(\frac{\partial \left<T(x,\tau)\right>}{\partial x} \right)^2 ~ \delta(x-x^{'}). \nonumber \\
\end{eqnarray}
The explicit solution is obtained by finding the spectral decomposition of the operator $~\frac{\partial ~}{\partial x} \kappa(x) \frac{\partial ~}{\partial x} ~$
\begin{equation}
\frac{\partial ~}{\partial x} \kappa(x) \frac{\partial ~}{\partial x} ~Q_k(x) ~ = ~ - k^2 ~ Q_k(x).
\end{equation}
 The details of the derivation are given in Appendix A. Here, we just quote the result for semi-infinite reservoirs in the limit $~ \epsilon ~= ~ \sqrt{\frac{\kappa_2}{\kappa_1}} ~\rightarrow~ 0 ~$
\begin{equation}
Q_k(x) ~\approx ~ \frac{b_k}{\sqrt{\kappa_2}} ~ \sin{\frac{kx}{\sqrt{\kappa_2}}} ~ \theta(x) ~ \theta(L-x),
\end{equation}
with
\begin{equation}
b_k^2 ~ \approx ~ \frac{\kappa_2}{L} ~ \sum_{n=-\infty}^{\infty} ~ \delta(k-n \frac{\pi \sqrt{\kappa_2}}{L})
\end{equation}
Hence, in the limit $~ \epsilon \rightarrow 0~$ the operator $~\frac{\partial ~}{\partial x} \kappa(x) \frac{\partial ~}{\partial x} ~$ has the same spectral decomposition as the operator ~$ {\kappa_2 \frac{\partial^2~}{\partial x^2}}~$ with Dirichlet boundary conditions at $~x=0,L $.
Assuming the system is in  the quasi-NESS described at the beginning of this section (i.e. ~$ \left<T(x,t) \right> ~= ~ T_s(x) ~= ~ T_1~ \theta(-x) ~+~ (T_1 + \beta x) ~ \theta(x) ~ \theta(L-x) ~ + ~ T_3 ~ \theta(x-L) ~$), the long-range contribution to the pair correlation function is
\begin{equation}
C^{LR}(x,x^{'};t) ~ = ~ \frac{2 }{L\pi^2} ~ \beta^2 ~\frac{k_B}{C_p} ~ \sum_{n=1}^{\infty} \frac{1}{n^2} \left( 1~- ~ e^{ -2 n^2 \frac{\kappa_2 \pi^2}{L^2} t } \right)~ \sin{n\frac{\pi x}{L}} ~ \sin{n \frac{\pi x^{'}}{L}},
\end{equation}
in the limit $~ \epsilon ~= ~ \sqrt{\frac{\kappa_2}{\kappa_1}} ~\rightarrow~ 0 ~$ and for times $~ t ~\ll~ \frac{1}{\epsilon^2} \frac{L}{2\pi^2\kappa_2} $. In the time regime during which the quasi-NESS is established $~   (\frac{L}{2\pi^2\kappa_2} ~ \ll~ t~ \ll~ \frac{1}{\epsilon^2} \frac{L}{2\pi^2\kappa_2})~$ the  long-range steady state correlations are of the form
\begin{eqnarray}
C^{LRSS}(x,x^{'}) & = & \frac{2 }{L\pi^2} ~ \beta^2 ~\frac{k_B}{C_p} ~ \sum_{n=1}^{\infty} \frac{1}{n^2} ~ \sin{n\frac{\pi x}{L}} ~ \sin{n \frac{\pi x^{'}}{L}} ~  ~ \nonumber \\
& = & \frac{k_B}{C_p} ~ \frac{\beta^2}{L} ~ \left\{ (L-x^{'}) ~x ~ \theta(x-x^{'}) ~ + ~ (L-x) ~x^{'} ~ \theta(x^{'}-x) \right\},
\end{eqnarray}
to lowest order in $~ \epsilon$. The correlations are proportional to the square of the temperature gradient  and  they decay linearly over a distance comparable to the system size. \\
\indent It is important to stress at this point that different boundary conditions from the ones described would give rise to long-range correlations of different form. However, if we impose the gradient by contact with macroscopic reservoirs, the appropriate boundary conditions are that  the fluctuations have a local equilibrium form at the edges of the system. 
\section{Diffusive lattice gas automaton.}
We construct a model system with stochastic microdynamics which, at the mesoscopic level, can be approximately described by the phenomenological theory  presented in the previous section. The model belongs to the class of lattice gas automata, where particles move on a  regular array at discrete time intervals. Since we are interested in the transport of a scalar quantity,  it is sufficient in two dimensions to consider the case of a square lattice  to obtain a diffusion equation at the macroscopic level.\\
\indent The  system analyzed is composed of particles moving on a two dimensional square lattice $~ {\cal{L}} ~$ of dimensions $~ (L_x+1) \times L_y ~$ \cite{foot0}. There are four channels per lattice node. Each channel corresponds to the particle velocity pointing towards one of the four nearest neighbors (~$i ~=~0~(right), ~1~ (up),~2~ (left), ~3~ (down) ~$). Channel $~i~$ at  node ~${\bf r} \in {\cal{L}}~$ is occupied if there is a particle at $~{\bf r}~$ with  velocity $~{\bf c_i}$. At  time $~t$, the state of the automaton is thus described by a set of bits 
\begin{equation}
\left\{~ n_i({\bf r},t); ~ ~ {\bf r} \in {\cal{L}}, \ \ i ~= ~ 0,1,2,3 \right\},
\end{equation}
where $~n_i({\bf r},t) ~$ is equal to  1 (0) if  channel  $~i~$ of node $~{\bf r}~$ is occupied (empty). There is an exclusion principle, which is the equivalent in the lattice gas of a hard-core potential preventing two particles from simultaneously occupying the same channel of a given node.\\
\indent The microscopic evolution of the automaton takes place in two stages. There is a propagation step, in which each particle in the automaton moves to the same channel of a neighboring node according to its velocity (see Fig. 2(a) ). For instance, if there is a particle in the channel ``up'' of node $~ {\bf r}$, the propagation step will move it to the channel ``up'' of the node $~ {\bf r}~+ {\bf \hat{y}} $, where $~{\bf \hat{y}}~$ is a unit vector  in the $~Y~$ direction. \\

\indent The second stage of the dynamics can be viewed as a ``collision'' step consisting of a random redistribution of the particles at each node (see Fig. 2(b) ) such that configurations compatible with the values of the conserved quantities (in our case,  the particle number only) are  equally probable outcomes of the collision step. This choice for the collision rules greatly simplifies the subsequent analytical derivation of the equations describing the macro- and mesoscopic evolution of the automaton.\\
\begin{figure}[h]
\[\psfig{figure=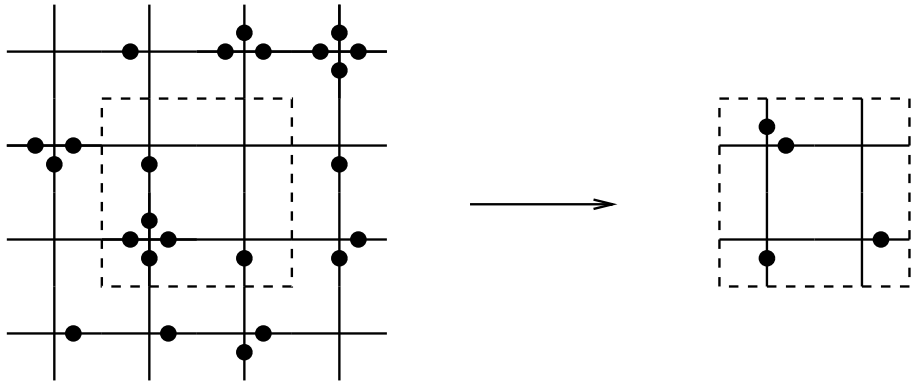,width=12cm,height=5cm}\]
\caption{\hspace{-1mm}(a)  Propagation step in a portion of the automaton. The motion takes place on a square lattice with  spacings $ \Delta x = \Delta y = 1$.  A dot on a link, close to node ${\bf r}$, indicates that there is a particle occupying that node with a unit velocity along the direction indicated by the link. 
} 
\end{figure} 
\setcounter{figure}{1}
\begin{figure}[h]
\[\psfig{figure=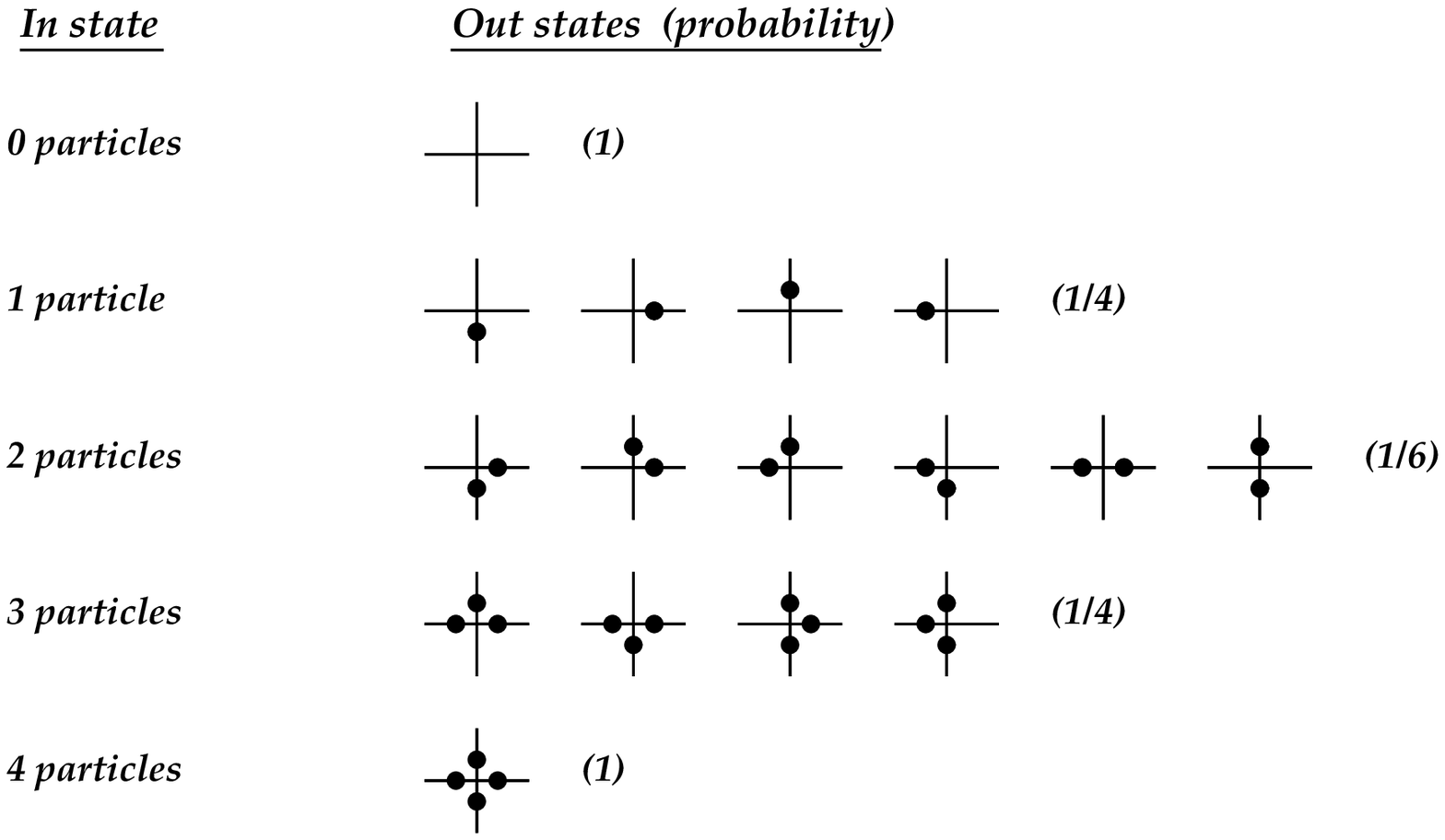,width=12cm,height=7cm}\]
\caption{\hspace{-1mm}(b) Collision step. The {\it in} state with $ ~m~=~0,..., 4 ~$ on a node yields any of the corresponding states with the same number of particles with probability $~\frac{1}{\nu_{\{s\}}}$, where $~\nu_{\{s\}}~$ is the multiplicity of the state (i.e. the number of compatible configurations).}

\end{figure} 

\indent From this set of rules that govern the dynamics of the automaton we can derive a hierarchy of contracted descriptions  containing progressively smaller  amounts of information about the details of the ``microscopic'' configuration of the automaton. The coarsest description is obtained by averaging the equation for the evolution of the channel occupation number  over the stochastic  dynamics. For the class of automata investigated this procedure yields  an exact  Boltzmann equation, which, in the hydrodynamic limit leads to  a diffusion equation. Section III.A contains a detailed discussion of this ``macroscopic'' level of description. A more refined (mesoscopic) description  is obtained  by incorporating the statistical properties of fluctuations. In section III.B, we investigate the behavior of static two point correlations of the density fluctuations. Following the arguments presented in section II, non-equilibrium constraints are implemented by randomly initializing the nodes at the automaton edge
s which  are  in contact with particle reservoirs. This procedure corresponds to assuming a local equilibrium form for the fluctuations at the boundaries. In the bulk, the fluctuations are seen to exhibit correlations that decay only algebraically over distances comparable to the  size of the system. These long-range correlations are very sensitive to different implementations of the non-equilibrium constraints. This point is illustrated in section III.C, where the constraints  are imposed  by direct manipulation of the  microscopic  configuration of the automaton. Simulations show that the qualitative features of the long-range correlations are modified, even though the macroscopic density profile is the same in both cases.

\subsection{Macroscopic dynamics: Diffusion equation.}
The starting point for the  derivation of  an equation for the time evolution of the average occupation  of a given node is the equation that describes the microdynamics of the automaton   
\begin{equation}
n_i({\bf r} + {\bf c_i},t+1) ~ = ~ {\cal{R}}(N({\bf r},t))
\end{equation}
where $~ N({\bf r},t) ~=~ \sum_{i=0}^3 n_i({\bf r},t)  ~$ is the occupation number of node $~{\bf r}~$ at time $~t$, and $~{\cal{R}}(N)~$ is a random function, which takes the value $~0~$ with probability $~ \left( 1- \frac{N}{4} \right) ~$ or the value $~1~$ with probability  $~\frac{N}{4} $, where N is an integer between $0$ and $4$. The Boltzmann equation is obtained by averaging Eq. (26) over the stochastic dynamics
\begin{equation}
f_i({\bf r},t+1) ~ = ~ \frac{1}{4} ~ \sum_{j=0}^3 ~ f_j({\bf r}- {\bf c_i},t),
\end{equation}
with the definition $~ f_i({\bf r},t) ~=~ \left< n_i({\bf r},t) \right> $. The angular brackets denote the average over the stochastic dynamics. Summing Eq. (27) over the index $i$, we obtain the equation governing the evolution of $~ \rho({\bf r},t) ~ = ~ \left<N({\bf r},t) \right> ~ = ~ \sum_{i=0}^3 f_i({\bf r},t)$, the local density (or average node occupation number)
\begin{equation}
\rho({\bf r},t+1) ~ = ~ \frac{1}{4} ~ \sum_{i=0}^3 ~ \rho( {\bf r}- {\bf c_i},t).
\end{equation}
Note that this is an {\it exact} Boltzmann equation  obtained without appealing to the {\it molecular chaos} hypothesis. The stochastic dynamics contains intrinsically the factorization of the occupation densities through the random redistribution of the  channel occupations \cite{boon95}. Equation (28) can be rewritten in the form
\begin{equation}
\rho({\bf r},t+1) ~-~ \rho({\bf r},t) ~ = ~ \frac{1}{4} ~ \sum_{i=0}^3 ~ \left( e^{- {\bf c_i} \cdot {\mbox{\boldmath $\nabla$}}   } -1 \right) \rho( {\bf r},t).
\end{equation}
In the continuous time limit and expanding to second order in the gradients (hydrodynamic limit)  Eq. (29)  yields the diffusion equation
\begin{equation}
\frac{\partial ~ }{\partial t} ~ \rho({\bf r},t) ~ = ~ D ~ \nabla_{{\bf r}}^2 ~ \rho({\bf r},t), 
\end{equation}
with a diffusion constant $~D = \frac{1}{4} $. In the derivation of Eq. (30) we have made use of the symmetries of the lattice, namely 
\begin{equation}
 \sum_{i=0}^3 {\bf c_i} ~ = ~ 0 ; ~ ~ \sum_{i=0}^3 {\bf c_i} {\bf c_i} ~ = ~ 2 \stackrel{\leftrightarrow}{1}, 
\end{equation}
where $~ \stackrel{\leftrightarrow}{1} ~$ is the unit tensor.\\
\indent For the sake of comparison of theoretical results with simulation data, it is convenient to define reduced quantities, which depend only on one of the spatial directions, $~X$, the direction along which the gradient is imposed, and on time. This is achieved by averaging over the remaining spatial direction, $~Y$. Along this direction periodic boundary conditions are imposed by identifying  nodes at $~y =0~$ and $~ y =  L_y $. Hence, the reduced density per node is
\begin{equation}
\rho(x,t) ~= ~ \frac{1}{L_y} ~ \sum_{y=0}^{L_y-1} ~ \rho({\bf r},t).
\end{equation}
In Fourier space, one can easily write  the time evolution 
\begin{eqnarray}
\delta \hat{\rho}(k,t) ~ =~  \delta\hat{\rho}(k,0)  ~ \exp{ \{- D k^2 t \}}, & & \\
&& k ~= ~ \frac{2\pi}{L_x} ~ n, ~ ~ n =0,1,2,... \nonumber
\end{eqnarray}
where $~\delta\hat{\rho}(k,t) ~$ is the Fourier transform of $~ \delta\rho(x,t) ~= ~ \rho(x,t) ~ - ~ \rho_s(x)$, the deviation of the actual density profile from  the asymptotic stationary profile $~\rho_s(x)$. This stationary profile is  a solution of the differential equation $~ \frac{\partial^2 ~}{\partial x^2} \rho_s(x)  =  0 $, and, depending on the boundary conditions imposed along $~X$, it can be a homogeneous equilibrium profile $~ \rho_s(x)  =  \rho_{eq} $, when $~ \rho(0) =  \rho(L_x)  =  \rho_{eq} $, or a non-equilibrium steady state, $~\rho_s(x)  ~ =~  \rho_s(0) ~+ ~ \beta ~ x; ~ ~ \beta ~= ~ \frac{\rho(L_x)-\rho(0)}{L_x} $, when $~ \rho(0) ~ \neq ~ \rho(L_x) $.\\
\indent The simulations of the automaton dynamics have been carried out under both equilibrium and non-equilibrium conditions. Figure 3(a) shows the time  evolution of the reduced density profile in a  $~ \left(  256 ~ \times ~2048\right) ~$ automaton with periodic boundary conditions in both the $X$ and $Y$ directions.  Initially the automaton exhibits  a square density profile, in which all channels of all nodes are occupied for $~x < 26$. This initial non-equilibrium profile evolves to a homogeneous equilibrium state characterized by a density per channel $~d_{eq} =  0.0976 ~$ [Note: The density per channel is  just one fourth of the density per node: $ ~d(x,t) ~ = ~ \frac{\rho(x,t)}{4} ~$]. \\
\indent The evolution of the different Fourier components of the density profile $~\rho(k,t)~$ is shown in Fig. 3(b). The amplitude of the mode $~ k=0 ~$ is constant, reflecting the global conservation of the number of particles. The remaining  modes, with $~ k \neq 0 $, have an average exponential decay that agrees with expression (33). From this exponential decay we can measure the diffusion constant, which has an experimental value of  $~D ~= ~0.252 ~ \pm ~0.002~$ \cite{foot1}, in excellent agreement with the value $~ D =  \frac{1}{4} $, predicted by the theory. The deviation arises from the terms neglected in approximating a finite difference equation (Eq. (28)) with a differential equation in both space and time (Eq. (30)).
\newpage
\vspace*{3cm}
\begin{figure}[h]
\[\psfig{figure=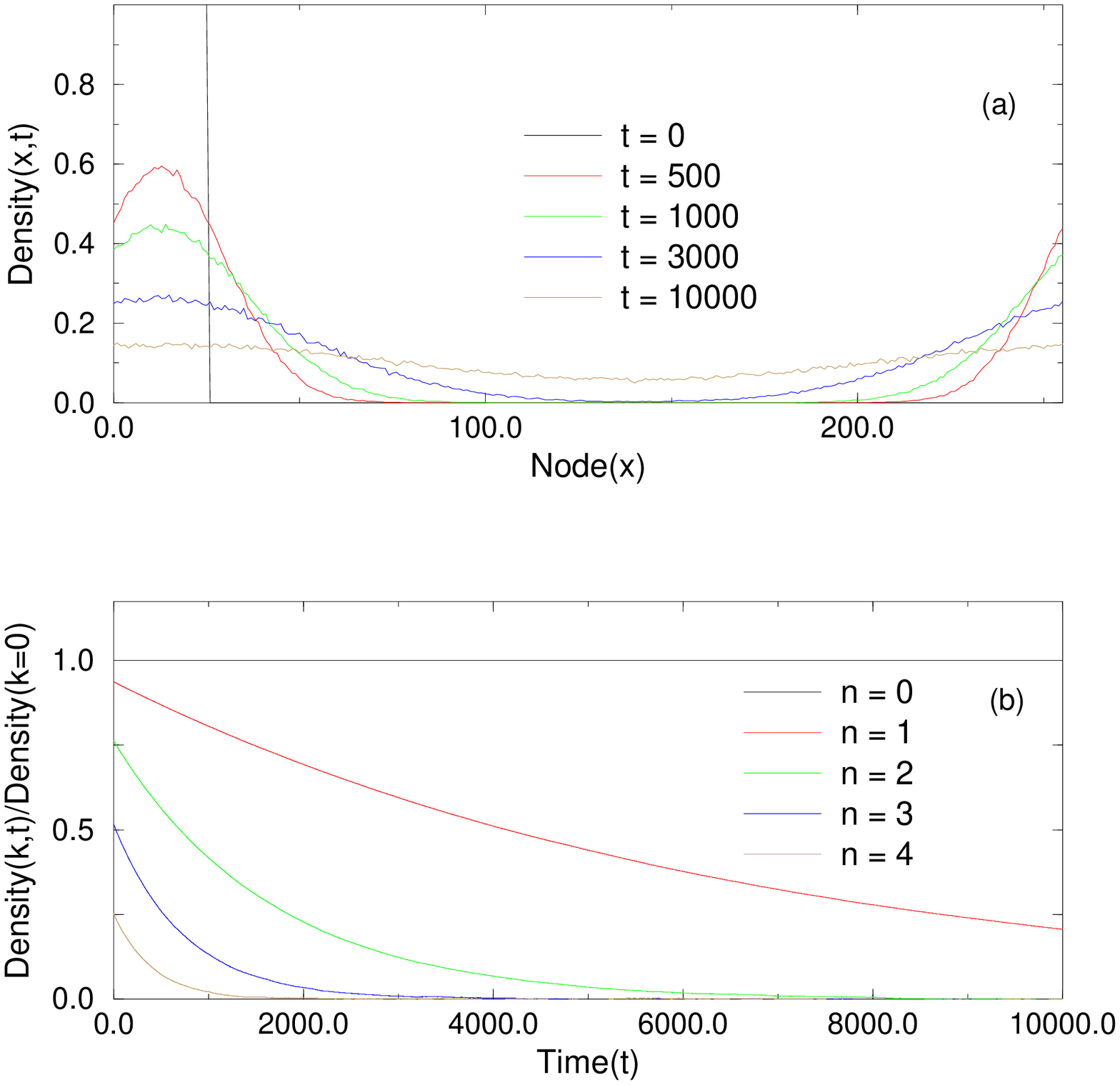,width=12cm,height=14cm}\]
\caption{Simulation data for a $~256 \times 2048 ~$ automaton with periodic boundary conditions in the $~X~$ and $~Y~$ directions. (a) Evolution of the reduced density profile along the $~X~$ direction. Initially the automaton presents a square density profile with all nodes with $~x<26~$ completely occupied. Asymptotically the automaton evolves to an equilibrium state characterized by a constant density profile. (b) Time evolution of the different Fourier components of the reduced density profile. In this and subsequent figures
time is given in automaton time step units, and space in units 
of lattice spacing.}
\end{figure} 

\indent The results of simulation under  non-equilibrium conditions in a somewhat smaller automaton ($256 \times 256$) are shown in Fig. 4. The initial state of the automaton is a square profile where the nodes on the left half of the lattice are fully occupied. The evolution quickly renders the profile smooth and eventually leads to a linear density profile, which is the asymptotic stationary state for the implemented boundary conditions ($d(x=0) = 1,  \ \ d(x=L_x)  =  0$). It is clear from Figs. 3(a) and 4  that the fluctuations in the $ ~256 \times 256~$ automaton are noticeably larger than in the $~ 256 \times 2048 ~$ automaton, in accordance with the idea that the relative size of fluctuations decreases with increasing system size.
\begin{figure}[h]
\[\psfig{figure=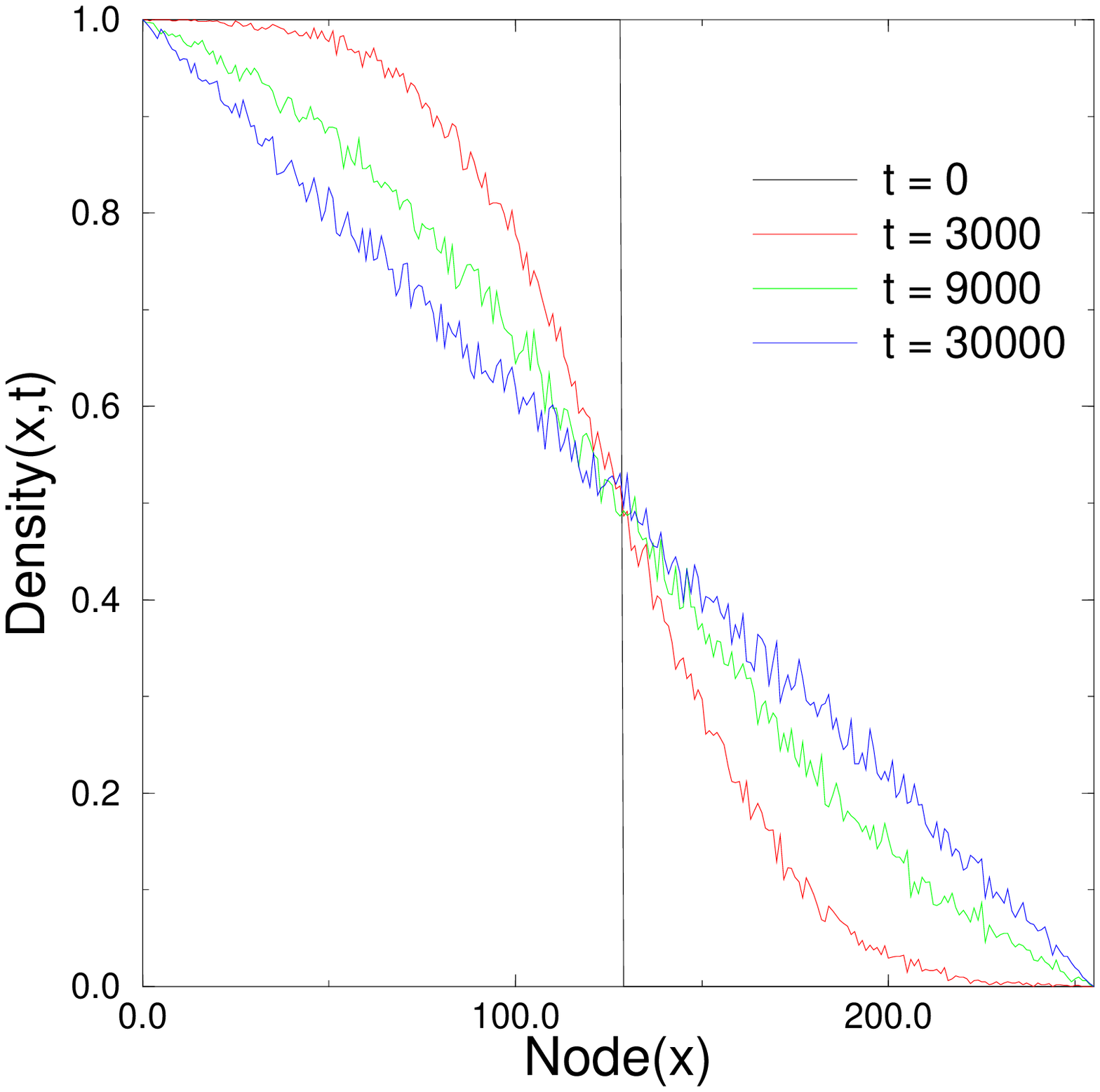,width=12cm,height=8cm}\]
\caption{Time evolution of the reduced density profile along the $X$ direction  in a $~256 \times 256~$ automaton under non-equilibrium conditions. Initially the system has a square profile with all the nodes for $~ 0 \le x<128 ~$ fully occupied and all other nodes $~ 128 \le x \le 255 ~$  empty. Asymptotically the system evolves to a steady state characterized by a linear density profile. Periodic boundary conditions are used for the $Y$ direction. For $X$, the density at the boundaries is kept constant at the values $~d(x=0)=1~$ and$~d(x=255) = 0$.}

\end{figure} 

\subsection{Mesoscopic dynamics: Fluctuations.}
We have derived an exact  Boltzmann equation, Eq. (28), which describes the dynamics of the single particle distribution of the automaton.  In order to characterize the fluctuations it is necessary to investigate the evolution of quantities associated with several particles. In particular, for two particles, we can formulate the equation
\begin{eqnarray}
&& \hspace{-0.7cm} n_i({\bf r}+ {\bf c_i}, t+1) ~  n_j({\bf r^{'}}+ {\bf c_j}, t+1) ~~   \nonumber \\
&& \hspace{-0.7cm}= ~ (1- \delta( {\bf r}, {\bf r^{'}}))  \sum_{ \{ s \},\{\sigma\}} \sum_{\{s^{'}\}, \{\sigma^{'} \}}  \sigma_i  \sigma_j^{'} \xi_{ \{ s \} \rightarrow \{\sigma\}} \xi_{ \{ s^{'} \} \rightarrow \{\sigma^{'}\}}  \delta(\{s\},\{n({\bf r},t)\}) \delta(\{s^{'}\},\{n({\bf r^{'}},t)\}) ~ \nonumber \\
&& \hspace{5.3cm}  + \ \delta({\bf r},{\bf r^{'}}) ~ \sum_{ \{ s \},\{\sigma\}} ~ \sigma_i  \sigma_j ~ \xi_{ \{ s \} \rightarrow \{\sigma\}} ~ \delta(\{s\},\{n({\bf r},t)\}).
\end{eqnarray}
The summations are over possible configurations of a node, with the notation $~ \{s\} \equiv \{s_i; i=0,1,2,3\} $. The random function $~ \xi_{ \{ s \} \rightarrow \{\sigma\}} ~$ is $1$ or $0$ depending on whether the  collision produces the configuration $~\{\sigma\} ~$ from the initial configuration $~ \{s\} ~$  or not. Upon averaging over the stochastic dynamics this quantity becomes  $~ A_{ \{ s \} \rightarrow \{\sigma\}} $, a matrix which has a block-diagonal form, provided that the configurations $~\{s\},\{\sigma\}~$ are grouped into equivalence classes according to the value of the conserved quantities. For the automaton considered here
\begin{equation}
  A_{ \{ s \} \rightarrow \{\sigma\}} ~ = ~ \frac{1}{\nu_{\{s\}}} ~ \delta(\sum_k s_k, \sum_k \sigma_k),
\end{equation}
where $~ \nu_{\{s\}} ~$ is the number of elements in the equivalence class to which the configuration $~ \{s\} ~$ belongs. Its value for the present case (see Appendix B) is
\begin{equation}
 \nu_{\{s\}} ~= ~ \left( \begin{array}{c} 4 \\ \sum_{k=0}^3 s_k \end{array} \right).
\end{equation}
Appendix B also contains a derivation of the result
\begin{equation}
\sum_{\{\sigma\}} ~\sigma_i ~ A_{ \{ s \} \rightarrow \{\sigma\}} ~ = ~ \frac{1}{4} ~ \sum_{k=0}^3 s_k,
\end{equation}
which is needed, together with the Boolean nature of the variables ($~\sigma_i^2 ~ = ~ \sigma_i~ $) to evaluate  the average of Eq. (34) over the stochastic dynamics:
\begin{eqnarray}
&&\left<n_i({\bf r}+ {\bf c_i}, t+1) ~ n_j({\bf r^{'}}+ {\bf c_j}, t+1) \right> ~= ~ \nonumber \\
&& ~ = ~ (1- \delta({\bf r},{\bf r^{'}})) ~ \frac{1}{16} \sum_{k,l} \left<n_k({\bf r},t) ~n_l({\bf r^{'}},t) \right> \nonumber \\
&& ~  + ~ \delta({\bf r}, {\bf r^{'}}) ~ \left[ \frac{1}{4} \delta_{ij} ~ \sum_k \left< n_k({\bf r},t) \right> ~+~ \frac{1}{12} ~ (1-\delta_{ij}) ~ \sum_{k,l} \left< n_k({\bf r},t)^{~} (n_l({\bf r},t)-1)  \right> \right] ~  ~ \nonumber \\
&& ~ = ~  \frac{1}{16} ~\left<N({\bf r},t) N({\bf r^{'}},t) \right> ~ + ~ \delta({\bf r},{\bf r^{'}}) \frac{1}{48} ~ (1-4\delta_{ij}) ~ \left[ \left<N^2({\bf r},t) \right> - 4 \left< N({\bf r},t) \right> \right].
\end{eqnarray}
Summing over the indices $i,j$ and making use of the equation for $~ \rho({\bf r},t) ~ = ~ \left< N({\bf r},t) \right> ~$ (Eq. (28)), we can derive the equation for the correlation function of the density fluctuations $~ C({\bf r},{\bf r^{'}};t) ~ = ~ \left<N({\bf r},t) N({\bf r^{'}},t) \right>  ~ - ~  \left<  N({\bf r},t)^{~} \right> \left< N({\bf r^{'}},t) \right> $,
\begin{eqnarray}
C({\bf r},{\bf r^{'}}; t+1) &-& C({\bf r},{\bf r^{'}},t) ~ = ~ \frac{1}{16}~ \sum_{ij} \left( e^{- {\bf c_i} \cdot {\mbox{\boldmath $\nabla$}}_{{\! \! \bf r}}} ~  e^{- {\bf c_j} \cdot {\mbox{\boldmath $\nabla$}}_{{\! \! \bf r^{'}}}} ~- ~1 \right) ~ C({\bf r},{\bf r^{'}};t) ~  \nonumber \\
&+&  \frac{1}{48} ~ \sum_{ij} e^{- {\bf c_i} \cdot {\mbox{\boldmath $\nabla$}}_{{\! \! \bf r}}} ~  e^{- {\bf c_j} \cdot {\mbox{\boldmath $\nabla$}}_{{\! \! \bf r^{'}}}} (1 ~ - ~ 4 ~ \delta_{ij}) \left[ \left<N^2({\bf r},t)\right> ~ - ~ 4 ~ \rho({\bf r},t) \right] \delta({\bf r},{\bf r^{'}}). \nonumber \\
\end{eqnarray}
\indent This exact equation can be approximated in the hydrodynamic limit (small gradients) and in the continuous time limit  by the following equation
\begin{eqnarray} 
\frac{\partial ~}{\partial t} C({\bf r},{\bf r^{'}};t) & = & D ~ \left( \nabla_{{\bf r}}^2 ~ +~ \nabla_{{\bf r^{'}}}^2 \right) ~ C({\bf r}, {\bf r^{'}};t) ~  \nonumber \\
 & & \ \ \ \ \ \ \ \  ~ - ~  {\mbox{\boldmath $\nabla$}}_{{\! \! \bf r}} \cdot  {\mbox{\boldmath $\nabla$}}_{{\! \! \bf r^{'}}}  \left[ \frac{2D}{3} \left(\left< N^2({\bf r},t) \right> ~ - ~ 4 ~ \rho({\bf r},t) \right)~ \delta({\bf r},{\bf r^{'}}) \right].
\end{eqnarray}
At this point, it is interesting to compare expression (40) with the theory of fluctuating hydrodynamics. Following the prescription given by Landau and Lifshitz \cite{landau}, we assume that $~ \delta N({\bf r},t) ~ = ~ N({\bf r},t) ~- ~ \rho({\bf r},t) ~$ is a random variable obeying a stochastic equation, which is constructed by adding a noise term to the macroscopic diffusion equation that governs the evolution of $~\rho({\bf r},t) ~ = ~ \left<N({\bf r},t) \right> ~$
\begin{equation}
\frac{\partial ~}{\partial t} \delta N({\bf r},t) ~ = ~  D ~  \nabla^2 ~  \delta N({\bf r},t) ~ + ~ {\mbox{\boldmath $\nabla$}} \cdot {\bf g}({\bf r},t).
\end{equation}
The term $~ {\bf g}({\bf r},t) ~$ is a random particle flux assumed to be Gaussian white noise with the properties
\begin{eqnarray}
\left<{\bf g}({\bf r},t) \right> & = & 0 \nonumber \\
\left<{\bf g}({\bf r},t) ~ {\bf g}({\bf r^{'}},t^{'}) \right> & = &  \stackrel{\leftrightarrow}{1}  ~ F_{FH}({\bf r},t) ~ \delta({\bf r},{\bf r^{'}}) ~ \delta(t-t^{'}),
\end{eqnarray}
where $~F_{FH}({\bf r},t)~$ is the amplitude of the noise.
Given this specification for the noise term, we can derive the equation for the pair correlation function for the density fluctuations, $~ C({\bf r},{\bf r^{'}};t) ~ = ~ \left<\delta N({\bf r},t) ~ \delta N({\bf r^{'}},t) \right> $, 
\begin{equation} 
\frac{\partial ~}{\partial t} C({\bf r},{\bf r^{'}};t) ~ = ~ D ~ \left( \nabla_{{\bf r}}^2 ~ +~ \nabla_{{\bf r^{'}}}^2 \right) ~ C({\bf r},{\bf r^{'}};t) ~ + ~  {\mbox{\boldmath $\nabla$}}_{{\! \! \bf r}} \cdot  {\mbox{\boldmath $\nabla$}}_{{\! \! \bf r^{'}}}  \left[ ~ F_{FH}({\bf r},t)  ~ \delta({\bf r},{\bf r^{'}}) \right].
\end{equation}
Knowing that, at equilibrium (i.e. for a constant density profile $~ \rho({\bf r},t) ~ = ~ \rho ~$), the correlation function for the density fluctuations is exactly (see Eq. (3.26) of Ref. \cite{ernst90})
\begin{equation}
C^{eq}({\bf r},{\bf r^{'}}) ~ = ~ \frac{\rho}{4} ~ (4 ~- ~ \rho) ~ \delta({\bf r},{\bf r^{'}}),
\end{equation}
we conclude that the strength of the noise term must be
\begin{equation}
F_{FH}^{eq} ~ = ~ \frac{D}{2} ~ \rho ~ (4 ~-~\rho)
\end{equation}
in order that Eq. (43) give the correct answer for the equilibrium correlations. \\
\indent Away from equilibrium, we make use of the hypothesis that the noise term, which reflects the effect of dynamics at the microscopic scale, has a local equilibrium form
\begin{equation}
F_{FH}({\bf r},t) ~ = ~ \frac{D}{2} ~ \rho({\bf r},t) ~ (4 ~-~\rho({\bf r},t)).
\end{equation}
With this hypothesis Eq. (43) becomes
\begin{eqnarray} 
\frac{\partial ~}{\partial t} C({\bf r},{\bf r^{'}};t) & = & D ~ \left( \nabla_{{\bf r}}^2 ~ + ~ \nabla_{{\bf r^{'}}}^2 \right) ~ C({\bf r},{\bf r^{'}};t) ~ \nonumber \\
&& \ \ \ \ \ \ \  + ~  {\mbox{\boldmath $\nabla$}}_{{\! \! \bf r}} \cdot  {\mbox{\boldmath $\nabla$}}_{{\! \! \bf r^{'}}}  \left[ \frac{D}{2} \rho({\bf r},t) ~ (4 ~-~\rho({\bf r},t))~  ~ \delta({\bf r},{\bf r^{'}}) \right].
\end{eqnarray}
We observe that this equation, which has been constructed in a phenomenological fashion, can be obtained from Eq. (40), which was derived directly from the microscopic dynamics, if we assume that the on-site correlations have a local equilibrium form
\begin{equation}
\left<N^2({\bf r},t) \right> ~ \approx ~ \left<N^2({\bf r},t) \right>_{LE} ~ = ~ \frac{3}{4} ~ \rho^2({\bf r},t) + \rho({\bf r},t).
\end{equation}
This is a good approximation in the thermodynamic limit, since the term neglected is of the order of one over size of the system.\\
\indent In order to study in detail the nature of these correlations, we focus on a $~ (L_x+1)\times L_y ~$ automaton in a non-equilibrium steady state characterized by an average stationary density profile, which is linear in the $~ X~$ direction
\begin{equation}
\rho_{s}({\bf r}) ~ = ~ \rho_s(x=0) ~ + ~ \beta~ x,
\end{equation}
where ~$ \beta ~ = ~ \frac{1}{L_x} ~ (\rho(x=L_x) ~ - ~ \rho(x=0)).$ \\
\indent In the pair correlation function for the density fluctuations we can distinguish two contributions: A local equilibrium term and a part  containing the long-range correlations
\begin{equation}
C({\bf r},{\bf r^{'}};t) ~ = ~ C^{LESS}({\bf r},{\bf r^{'}}) ~ + ~ C^{LR}({\bf r},{\bf r^{'}};t),
\end{equation}
where
\begin{equation}
 C^{LESS}({\bf r},{\bf r^{'}}) ~ = ~ \frac{1}{4} ~ \rho_s({\bf r}) ~ \left( 4 ~ - ~ \rho_s({\bf r}) \right) ~ \delta({\bf r},{\bf r^{'}}).
\end{equation}
Substituting expressions (50-51) in Eq. (39), we obtain
\begin{eqnarray}
C^{LR}({\bf r},{\bf r^{'}}; t+1) &-& C^{LR}({\bf r},{\bf r^{'}},t) ~ = ~ \frac{1}{16}~ \sum_{ij} \left( e^{- {\bf c_i} \cdot {\mbox{\boldmath $\nabla$}}_{{\! \! \bf r}}} ~  e^{- {\bf c_j} \cdot {\mbox{\boldmath $\nabla$}}_{{\! \! \bf r^{'}}}} ~- ~1 \right) ~ C^{LR}({\bf r},{\bf r^{'}};t) ~  \nonumber \\
&+&  \frac{1}{48} ~ \sum_{ij} e^{- {\bf c_i} \cdot {\mbox{\boldmath $\nabla$}}_{{\! \! \bf r}}} ~  e^{- {\bf c_j} \cdot {\mbox{\boldmath $\nabla$}}_{{\! \! \bf r^{'}}}} (1 ~ - ~ 4 ~ \delta_{ij}) \left[ C^{LR}({\bf r},{\bf r};t) ~ \delta({\bf r},{\bf r^{'}}) \right] ~  \nonumber \\
&+& \frac{1}{4} ~ \sum_i \left( e^{- {\bf c_i} \cdot ({\mbox{\boldmath $\nabla$}}_{{\! \! \bf r}} + {\mbox{\boldmath $\nabla$}}_{{\! \! \bf r^{'}}})} -1 \right) ~ \left[  \frac{1}{4} ~ \rho_s({\bf r}) ~ \left( 4 ~ - ~ \rho_s({\bf r}) \right) \right] ~ \delta({\bf r}, {\bf r^{'}}).
\end{eqnarray}
Using  Eqs. (31) and (49) to simplify the last term, we have
\begin{eqnarray}
C^{LR}({\bf r},{\bf r^{'}}; t+1) &-& C^{LR}({\bf r},{\bf r^{'}},t) ~ = ~ \frac{1}{16}~ \sum_{ij} \left( e^{- {\bf c_i} \cdot {\mbox{\boldmath $\nabla$}}_{{\! \! \bf r}}} ~  e^{- {\bf c_j} \cdot {\mbox{\boldmath $\nabla$}}_{{\! \! \bf r^{'}}}} ~- ~1 \right) ~ C^{LR}({\bf r},{\bf r^{'}};t) ~  \nonumber \\
&+&  \frac{1}{48} ~ \sum_{ij} e^{- {\bf c_i} \cdot {\mbox{\boldmath $\nabla$}}_{{\! \! \bf r}}} ~  e^{- {\bf c_j} \cdot {\mbox{\boldmath $\nabla$}}_{{\! \! \bf r^{'}}}} (1 ~ - ~ 4 ~ \delta_{ij}) \left[ C^{LR}({\bf r},{\bf r};t) ~ \delta({\bf r},{\bf r^{'}}) \right] ~ \nonumber \\
&-& \frac{1}{8} ~ \beta^2 ~ \delta({\bf r}, {\bf r^{'}}).
\end{eqnarray} 
If we neglect the second term on the right-hand side of Eq. (53), which is equivalent to  assuming that the on-site correlations have a local equilibrium form (i.e. $~ \left<N^2({\bf r},t) \right> ~ \approx ~ \left<N^2({\bf r},t) \right>_{LE} ~$), we find that this equation  describes a random-walk in a four-dimensional cubic lattice $~({\bf r}, {\bf r^{'}}) ~$ with sources of constant strength along the surface  $~ {\bf r}  =  {\bf r^{'}} $. In order to visualize the mechanism by which the long-range correlations are built up, consider the evolution of a system initially at local equilibrium ($~ C^{LR}({\bf r},{\bf r^{'}}; t=0) ~ = ~ 0 ~$), with a stationary density profile given by Eq. (49). On average,  channel$~i~$ of node $~{\bf r} ~$ has an occupation $~ f_i({\bf r}) ~=~ \frac{1}{4} ~ \rho_s({\bf r}) ,$  for $~ i ~=~0,1,2,3$. After the propagation step, the average occupation per channel at node $~ {\bf r} ~$ is 
\begin{eqnarray*}
f_0({\bf r}) & = & \frac{1}{4} ~ (\rho_s({\bf r}) ~ - ~ \beta ~ \Delta x) \\
f_1({\bf r}) & = & \frac{1}{4} ~ \rho_s({\bf r}) \\
f_2({\bf r}) & = & \frac{1}{4} ~ (\rho_s({\bf r}) ~ + ~ \beta ~ \Delta x) \\
f_3({\bf r}) & = & \frac{1}{4} ~ \rho_s({\bf r}),
\end{eqnarray*}
where $~ \Delta x ~$ is the lattice spacing in the $X$ direction, which we have
taken to be equal to one throughout the paper.
The pair correlations are local and equal to $~ \left< \delta n_i({\bf r})~ \delta n_j({\bf r}) \right> ~ = ~ \delta_{ij} ~ f_i({\bf r}) ~ (1 ~- ~  f_i({\bf r})) $, which yields 
\begin{equation}
C({\bf r},{\bf r}) ~ = ~ \frac{1}{4} ~ \rho_s({\bf r}) ~ \left( 4 ~ - ~ \rho_s({\bf r}) \right) ~ - ~ \frac{1}{8} ~ \beta^2.
\end{equation}
After the collision step the particles are redistributed at random, which implies that, on average, the channel occupation number is $~ f_i({\bf r}) ~ = ~ \frac{1}{4} ~ \rho_s({\bf r}), $ for $~  i ~=~0,1,2,3 $, as initially. Since the number of particles at each node is conserved in the collision step, the total correlations (Eq. (54)) do not change, except that the non-equilibrium contribution is now completely off-diagonal in the channels:
\begin{equation}
C_{ij}({\bf r},{\bf r}) ~ =~ \left< \delta n_i({\bf r})~ \delta n_j({\bf r}) \right> ~ = ~ \delta_{ij} ~ f_i({\bf r}) ~ (1 ~- ~  f_i({\bf r}))  ~ - ~ (1-\delta_{ij}) \frac{1}{96} ~ \beta^2.
\end{equation}
In $~ C_{ij}({\bf r},{\bf r}) ~$ the diagonal terms $~(i=j)~$ have a local equilibrium form and  they remain local upon propagation. The off-diagonal terms ($~ i \neq j~$) give rise to the source term proportional to $~ - \frac{\beta^2}{8} ~$ that appears in Eq. (53). Upon propagation, these off-diagonal correlations become non-local and perform a random walk (diffusion in the continuous limit) in the four dimensional lattice $~({\bf r},{\bf r^{'}})$.  This accounts for the first term that appears on the right-hand side of Eq. (53). The only contribution unaccounted for is the second term in the r.h.s. of Eq. (53), which  corresponds to long-range correlations created from pre-existing long-range correlations. It is small in the thermodynamic limit, and we shall neglect it henceforth.\\ 
\indent Finally, after neglecting the on-site contributions from the long-range term, and in the hydrodynamic and continuous time limits, we approximate Eq. (53) by 
\begin{equation}
\frac{\partial ~ }{\partial t} ~ C^{LR}({\bf r},{\bf r^{'}};t) ~= ~ D ~ \left( \nabla^2_{{\bf r}} ~+~ \nabla^2_{{\bf r^{'}}} \right) C^{LR}({\bf r},{\bf r^{'}};t) ~ -  ~ \frac{1}{8} ~ \beta^2 ~ \delta({\bf r},{\bf r^{'}}).
\end{equation}
In order to compare with the simulations, we define reduced quantities by averaging over The $Y$ direction (orthogonal to the direction  of the gradient)
\begin{equation}
C(x,x^{'};t) ~ = ~  \sum_{y=0}^{L_y -1} ~\sum_{y^{'}=0}^{L_y -1} ~ C({\bf r},{\bf r^{'}};t). 
\end{equation}
We impose periodic boundary conditions in the $~Y~$ direction and  Dirichlet boundary conditions in the $~ X~$ direction. As discussed in the previous section, the Dirichlet boundary conditions (i.e. that the correlations at the edges of the system have a local equilibrium form), correspond to the system being in contact with two reservoirs of different chemical potential. This corresponds to the paradigm of the automaton implemented in our simulations, where the configuration of all the nodes at $~x=0,L_x~$ are reinitialized  randomly at  every time step, with the constraint that the average densities at $~x=0, L_x ~$ be constant (although, in general, different from each other). The asymptotic solution of (56) with these boundary conditions is \cite{nicolis84,garcia87}
\begin{equation}
 C^{NESS}(x,x^{'}) ~ = ~ C^{LESS}(x,x^{'}) ~ + ~ C^{LRSS}(x,x^{'}),
\end{equation}
with
\begin{equation}
C^{LESS}(x,x^{'}) ~ = ~ \frac{1}{4} ~ L_y ~ \rho_s(x) ~ (4 ~-~ \rho_s(x)) ~ \delta(x,x^{'}), 
\end{equation}
and
\begin{equation} 
C^{LRSS}(x,x^{'}) ~ = ~- \frac{\beta^2}{4} ~ \frac{L_y}{L_x} \left\{ x^{'}~ (L_x-x) ~ \theta(x-x^{'}) ~ + ~ x~ (L_x-x^{'}) ~ \theta(x^{'}-x) \right\}.  
\end{equation}  
\indent Simulations have been carried out in $~ 11 \times 11~$ automata with both equilibrium and non-equilibrium constraints. Figure 5 (a) shows the density fluctuation correlations  in a closed system with periodic boundary conditions in both the $X$ and $Y$ directions. It can be seen that together with the local-equilibrium term, there is a long-range contribution, which is constant and negative, as predicted by Bussemaker {\it et al} \cite{bussemaker95}. This long-range contribution is well-known in classical statistical mechanics \cite{macquarrie}  and appears solely as a consequence of  global conservation of particles (see Appendix C). \\
\indent  More physically relevant simulations are carried out in automata with periodic boundary conditions in the $Y$ direction and in contact with particle reservoirs at the edges perpendicular to the $X$ direction. Figure 5 (b) shows the density fluctuation correlations between the middle node and the other nodes $~C^{eq}(x,x^{'}=5)~$ in an equilibrium system with a homogeneous density profile. The  average density per channel in this automaton $~ d = 0.5$. Note that the correlations are local and exactly equal (up to sampling errors) to the  value $~L_y = 11~$ (see Eq. (59)). The average was performed over $~10^8~$ time steps.\\
\vspace*{-8cm}
\begin{figure}[h]
\[\psfig{figure=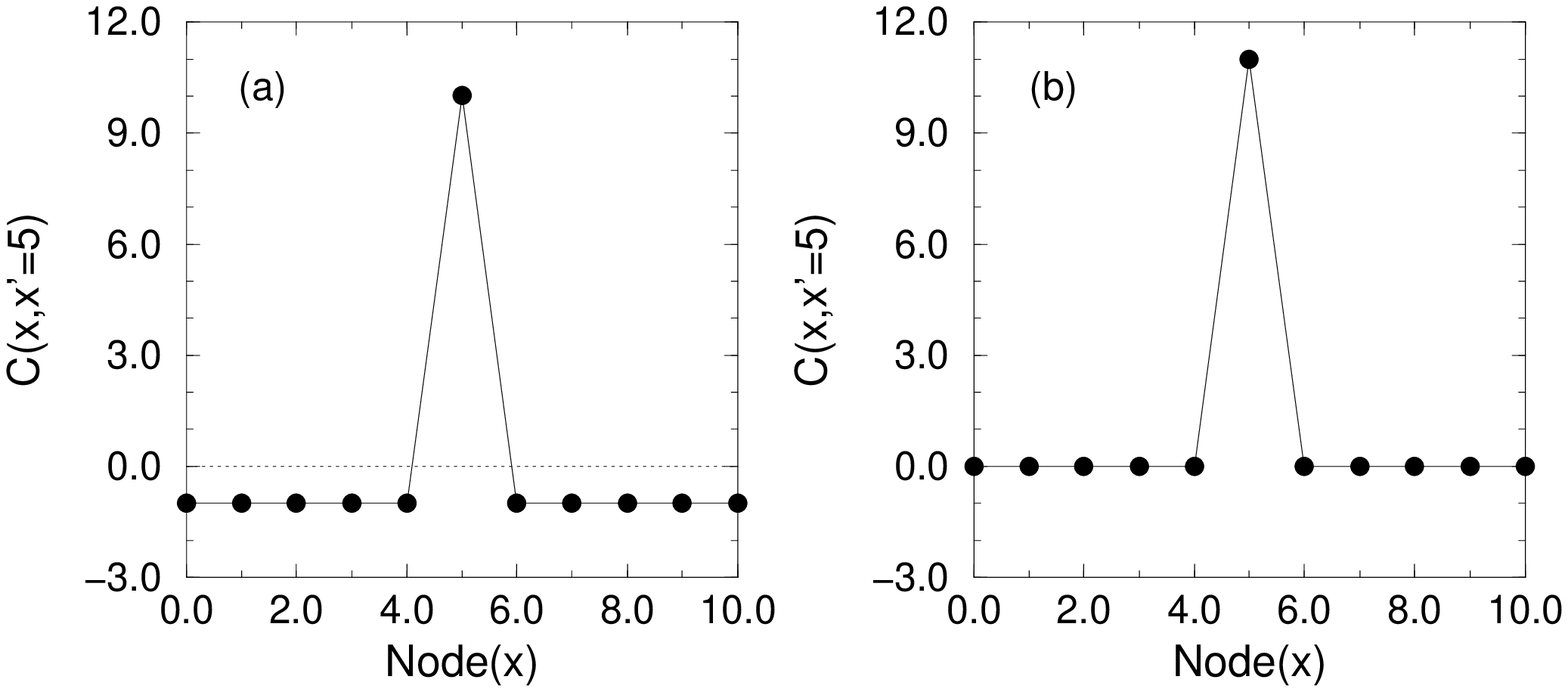,width=16cm,height=16cm}\]
\caption{ Equilibrium correlations of the density fluctuations, $~C^{eq}(x,x^{'}=5)$, in an $~ 11 \times 11 ~$ automaton. (a) Periodic boundary conditions in both  $X$ and $Y$ directions; the average density is $~d = 0.5$. (b) Periodic boundary conditions in the $Y$ direction; for the $~X~$ direction, the average density of the boundary nodes is kept fixed at the value $~d(x=0)=d(x=10) = 0.5$.}
\end{figure} 

\indent Figure 6 (a) contains a plot of  the  correlations between fluctuations at nodes  $~x=5~$ and nodes with arbitrary $x$, $ ~ C^{NESS}(x,x^{'}=5)$, for a system that is maintained in a non-equilibrium steady state by fixing the boundary densities at the values $~d(x=0)=1~$ and $~ d(x=10) =0 $. 
Figure 6(b) is a plot of the long-range component $~C^{LRSS}(x,x^{'}=5)$ of the non-equilibrium correlations depicted in Fig. 6(a). The full line is simulation data, obtained by averaging over $~10^8~$ time steps. The dotted line corresponds to the approximation given by fluctuating hydrodynamics, Eqs. (58-60). A similar comparison is shown in Fig. 6(c) for $~C^{LRSS}(x,x^{'}=3)$.\\
\indent Note that the equation for $~C^{LR}({\bf r},{\bf r^{'}};t) ~$ (Eq. (56)) contains information about the magnitude of the gradient ($\beta$), but not about its direction. This means that correlations are also long-ranged along the direction {\it perpendicular} to the concentration gradient. This is illustrated in Fig. 6(d), where we have plotted $~C(y,y'=5) ~= ~ \sum_{x,x^{'}} ~ C^{LRSS}(x,y,x^{'},5) ~$ for the same automaton. The correlations are different from those depicted in Fig. 6(b) owing to the asymmetry introduced by the type of boundary conditions (fixed in the $X$ direction, periodic in the $Y$ direction). The theoretical curve is obtained by  solving Eq. (56) by Fourier analysis. 
\vspace*{3cm}
\begin{figure}[h]
\[\hspace*{-3cm}\psfig{figure=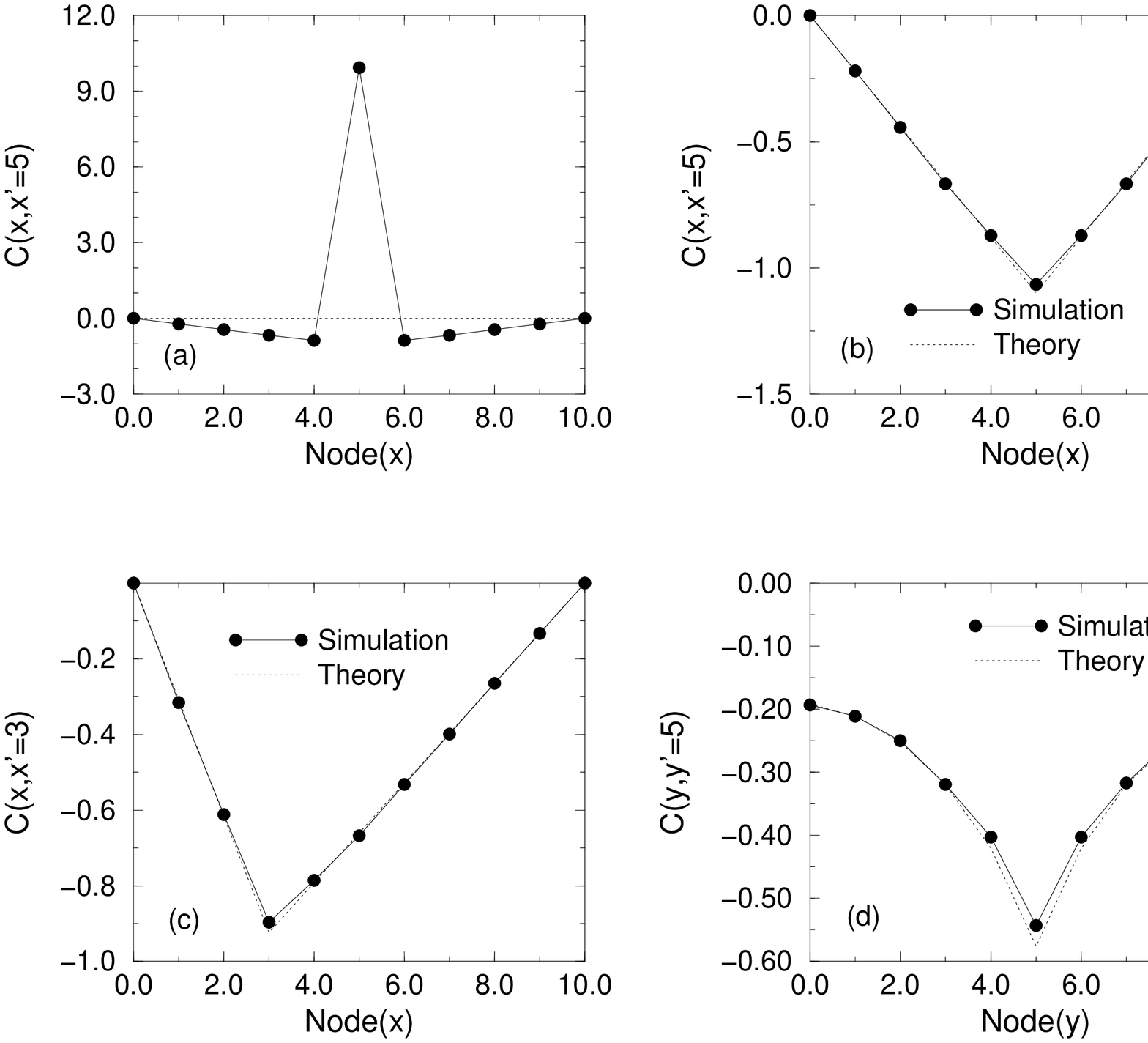,width=14cm,height=14cm}\]
\caption{Simulation data and theoretical results for an $~11 \times 11~$ automaton in a non-equilibrium steady state with $~d(x=0)=1$, $d(x=10) = 0 $. (a) Correlations of the density fluctuations,$~ C^{NESS}(x,x^{'}=5)$. (b) Long-range contribution  $~C^{LRSS}(x,x^{'}=5) $. (c) Long range contribution $~C^{LRSS}(x,x^{'}=3)$. (d) Long-range contribution $~C^{LRSS}(y,y^{'}=5)$.}
\end{figure} 

\subsection{ Effect of  boundary conditions.}
Unless we have access to its microscopic configuration, the only way to impose a non-equilibrium density profile in a diffusive system is to establish contact with  two particle  reservoirs of different chemical potential. Provided that the diffusion coefficient is independent of density, a system  under these constraints will display, on average, a linear  density profile.  Furthermore, as discussed in section II, the density fluctuations  exhibit both  local-equilibrium and  long-range correlations in the bulk, but only local-equilibrium correlations at the boundaries.\\
\indent One of the advantages of investigating  the type of problem under consideration with an automaton  is that we are able to manipulate the configuration in each node at  will. For instance, the system can be forced by injecting and withdrawing particles in such a way  that it asymptotically displays an arbitrary stationary density profile. Obviously, this manipulation involves an important modification in the nature of the fluctuations, which are extremely sensitive to how the microscopic configuration of the automaton is altered, and the loss of detailed balance in those nodes where the external perturbation is acting.  \\
\indent In order to illustrate the effect of different  boundary conditions on the long-range fluctuations, we have carried out simulations in a diffusive LGA, where for $~ 0 \le x \le L_x ~$ a stationary reduced density profile  of the form
\begin{equation}
\rho_s(x) ~ = ~ \rho_1 \theta(a-x) ~ +~  (\rho_1-\beta x) ~ \theta(x-a) \theta(b-x) ~ + ~ \rho_3 ~ \theta(x-b), 
\end{equation}
is established by creating particles at $ ~ x=a$, and annihilating particles at $~x=b~$ at  appropriate constant rates (see Fig. 7). This procedure  is very different from the previous implementation, in which  at every step all the nodes with $~ x = a, b ~$ are re-initialized at random, in such a way as to maintain the respective average densities constant. As far as fluctuations are concerned, this implied that the correlations between  fluctuations in nodes at the boundaries and  those in bulk nodes are systematically  destroyed at every time step. As a consequence, long-range correlations in such a system are confined in the region $~ a<x<b ~$ while the rest of the system ($~ 0 \le x \le a, ~ b \le x \le L_x~$) exhibits only local-equilibrium fluctuations. \\
\indent For the implementation with the creation/annihilation procedure, in the continuous-time and  hydrodynamic limit, the Landau equation governing the automaton evolution is  
\begin{equation}
\frac{\partial ~}{\partial t} \rho(x,t) ~ = ~ D \frac{\partial^2 ~}{\partial x^2} \rho(x,t) ~ + ~ \frac{\partial ~}{\partial x} g(x,t) + F_{ext}(x,t),
\end{equation}
where the diffusion constant is  $~ D = \frac{1}{4}$, the quantity $~ g(x,t) ~$ is the familiar random particle flux with local equilibrium form, and $~F_{ext}(x,t)~$ is an external field representing the effect of injecting and withdrawing particles at $~ x = a, b$.  The average of this external field is stationary and equal to 
\begin{equation}
{\bar{F}}_{ext}(x) ~ = ~ \left<F_{ext}(x,t) \right> ~ = ~ - ~D ~ \frac{\partial^2 ~}{\partial x^2} \rho_s(x) ~ + ~ \beta ~ D ~ \left[\delta(x-a) ~ - ~ \delta(x-b) \right]. 
\end{equation}
The external field has a non-vanishing fluctuating part,  which can be assumed to be Gaussian white noise provided that  we are only interested in calculating  pair correlations. Since we are manipulating the system only locally at $~ x= a,b$, the fluctuations of the external field will also be local
\begin{eqnarray}
\left< \left(F_{ext}(x,t) -\bar{F}_{ext}(x) \right) ~ \left(F_{ext}(y,t^{'}) - \bar{F}_{ext}(y) \right) \right> &  & \nonumber \\
&& \hspace{-7cm} = ~  D ~ \delta(t-t^{'}) ~\left[ \Delta_a ~ \delta(x-a) \delta(y-a) ~ + ~  \Delta_b ~ \delta(x-b) \delta(y-b) \right],
\end{eqnarray}
where $ ~ \Delta_a~ $ ($\Delta_b $) measures the intensity  of  particle creation  (annihilation).
The system is in contact with reservoirs at $~ x=0~$ and $~x=L_x~$ of appropriate chemical potential so that the average density is constant and equal to $~ \rho_1~$ and $~\rho_3$, respectively. \\
Again, the correlation function of the steady state density fluctuations can be separated into two terms
\begin{equation}
C^{NESS}(x,x^{'}) ~ = ~ C^{LESS}(x,x^{'}) ~ + ~ C^{LRSS}(x,x^{'}),
\end{equation}
with a local equilibrium contribution,
\begin{equation}
 C^{LESS}(x,x^{'}) ~ = ~ \frac{1}{4} ~ L_y ~ \rho_s(x) ~ (4 ~-~ \rho_s(x)) ~ \delta(x,x^{'}), 
\end{equation}
and a long-range contribution, which can be expressed as a double Fourier series 
\begin{equation}
C^{LRSS}(x,x^{'}) ~ = ~ \sum_{n=1}^{\infty} ~ \sum_{m=1}^{\infty} ~ C_{nm} ~ \sin{\frac{n\pi}{L_x}x} ~  \sin{\frac{m\pi}{L_x}x^{'}},
\end{equation}
with 
\begin{eqnarray*}
C_{nm} & = & \frac{1}{n^2+m^2} ~ \left(\frac{2}{\pi} \right)^2 \left[ - \beta^2 ~ \frac{L_y}{2} ~ \int_a^b ~ dz ~ \sin{\frac{n\pi}{L_x}z} ~  \sin{\frac{m\pi}{L_x}z} ~ \right. \\
& & \left. \ \ \ \ \ \ \ \ \ \ + ~ c_a ~  \sin{\frac{n\pi}{L_x}a} ~  \sin{\frac{n\pi}{L_x}a} ~ + ~ c_b~ \sin{\frac{n\pi}{L_x}b} ~  \sin{\frac{n\pi}{L_x}b} \right],
\end{eqnarray*}
where
\begin{eqnarray*}
c_a & = & \Delta_a ~-~ \frac{1}{4}~\beta~(4-2 \rho_1) \\
c_b & = & \Delta_a ~+~ \frac{1}{4}~\beta~(4-2 \rho_3).
\end{eqnarray*} 

\indent Simulations have been carried in a $~ 31 \times 31 ~$ automaton with the parameters $ ~ a=10, ~ b = 20, ~ d_1 = 0.75, d_3 = 0.25 $.  For these values, $ ~ c_a = c_b = c $, for symmetry reasons.  Figure 7 shows the  density profile  averaged over $~10^4~$ time-steps, which is indeed of the form given by Eq. (61). \\
\begin{figure}[h]
\[\psfig{figure=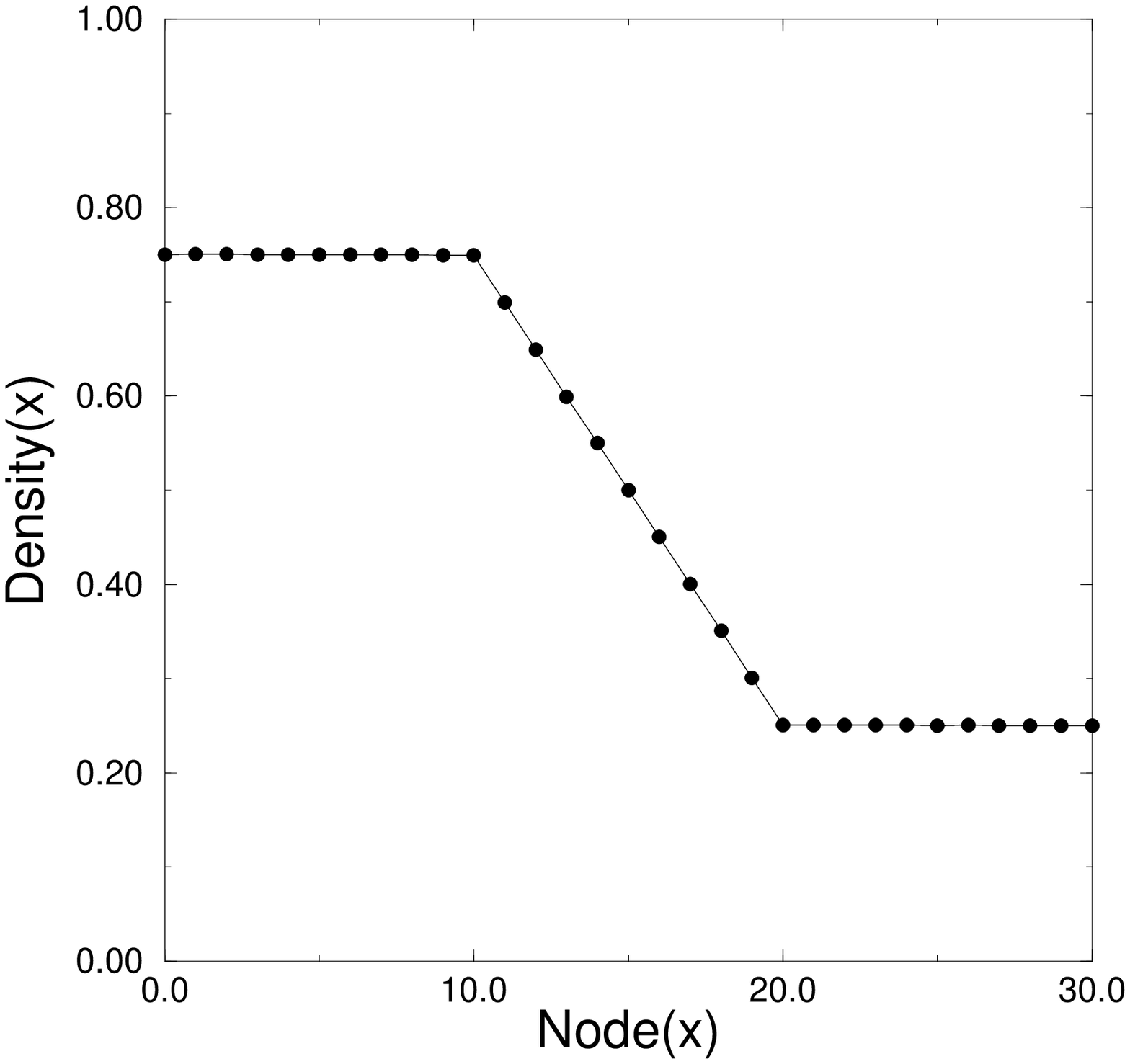,width=12cm,height=8cm}\]
\caption{Averaged density profile in a $~31\times 31~$ automaton, obtained by creating particles at $ ~ a=10$, and annihilating them at $~ b=20 ~$ at appropriate rates. The average densities at the boundaries  are kept at $~d_1= d(x=0) = 0.75~$ and $~ d_3 = d(x=30) = 0.25$.}
\end{figure} 
\vspace*{-8cm}
\begin{figure}[h]
\[\psfig{figure=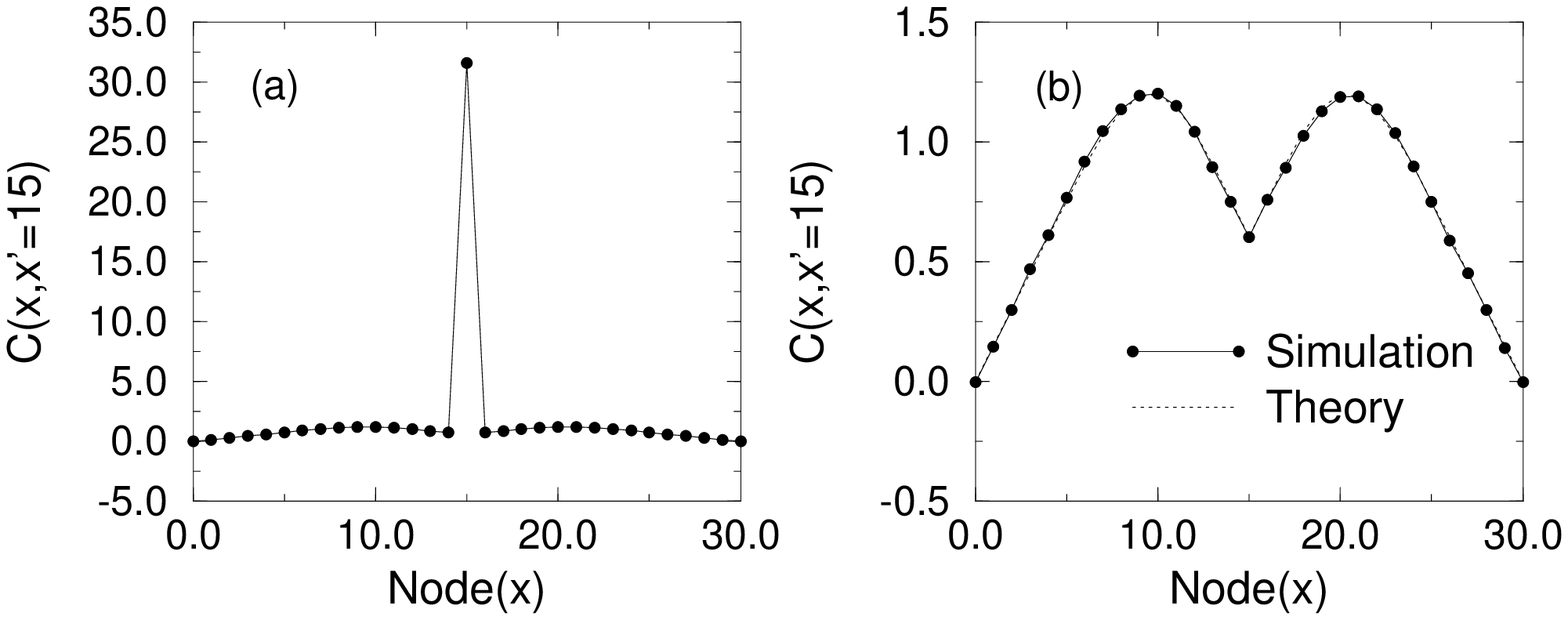,width=16cm,height=16cm}\]
\caption{ (a) Steady state correlations, $~C(x,x^{'}=15)$, in a $~31\times 31~$ automaton in  the non-equilibrium state depicted in Fig. 7. (b) Long-range contribution, $~C^{LRSS}(x,x^{'}=15)$.}
\end{figure}

\indent In Fig. 8(a), the density fluctuation correlations $~C^{NESS}(x,x^{'})~$ are plotted for $~x^{'}=15$. The continuous line in Fig. 8(b) connects the simulation data for the long-range correlations at steady state between an arbitrary point and the middle node  $~ C^{LRSS}(x,x^{'}=15)$, obtained after averaging over $~5 \times 10^7 ~$ time steps.  The dashed line represents the approximate solution, given by expression (67) (the Fourier series has been truncated after $~n,m = 1000~$), for a value  $~ c=9.1$, which has been chosen so as  to obtain the best fit to the simulation data. 
The agreement between simulations and the approximation given by Fluctuating Hydrodynamics  is very good. There are several features which call for  comments: First, the correlations extend beyond  the central part of the system $~ a<x<b $, where there is a concentration gradient: They have a  non-zero value for all $ ~ 0 < x < L_x $. Second, the form of the correlations is qualitatively different from those  obtained in a system with a linear density profile for $~ a \le x \le b $, maintained by contact  with particle reservoirs at the extremes. In particular they are  {\it positive}, non-monotonic as a function of distance  and furthermore they are non-linear. Nonetheless, the fluctuations in the ``bulk'' (i.e. around the central node)  are  qualitatively similar in both situations (compare Figs. 6(a) and 8(a), and Figs. 6(b) and 8(b)).
\section{Summary.}
We have constructed a lattice gas automaton with a set of fully stochastic collision rules, giving rise to diffusive behavior in the hydrodynamic and continuous time limit. This model system exhibits spontaneous fluctuations similar to those found in actual fluids \cite{grosfils92}.  The statistical properties of these fluctuations are the main objective of the present study. In the simulations  the automaton is kept at equilibrium or away from equilibrium by placing it in contact with particle reservoirs of equal or different chemical potentials, respectively. From a computational point of view, these conditions are implemented by randomly initializing the nodes at the boundaries of the  system at every time step. Under equilibrium constraints, the correlations of the density fluctuations are localized on each node, a result which is consistent with the observation that in fluids away from critical points or hydrodynamic instabilities, pair correlations decay on a microscopic lengthscale. The local nature o
f equilibrium correlations is the result of a balance between noise sources, which is lost as soon as  the system is removed from equilibrium. It is thus seen that in a non-equilibrium steady state  the density fluctuations  are correlated not only locally in an equilibrium-like manner, but also over large distances. This implies that   the non-equilibrium distribution function is not  factorized, in contrast to the equilibrium case. Both the local-equilibrium and the long-range contributions to   correlations have been measured in computer simulations. The results of  simulations  have  been compared to theoretical expressions derived from the stochastic microdynamics of the automaton. As a first approximation to these exact expressions, one can derive expressions obtained by means  of the phenomenological theory of fluctuating hydrodynamics, which  are in excellent agreement with the results of simulations even in fairly small automata.\\
\indent For  the  automata  analyzed in this paper where the microscopic evolution preserves the particle number and where a preferred spatial direction is established by the imposition of a density gradient,  global detailed balance is absent  as a consequence the imposition of non-equilibrium constraints at the boundaries of the system. Long-range correlations of the type described here appear generically in dynamical systems   with  conserved quantities,  having  large scale spatial anisotropy and lacking detailed balance.

\acknowledgements

It is a pleasure to acknowledge  Irwin Oppenheim for fruitful discussions. This work was supported by the EC under contract  ERBCHBG-CT93-0404. J. P. B. acknowledges support by the Fonds National de la Recherche Scientifique (FNRS, Belgium). P. G. is grateful to the Japan Society for the Promotion of Science for their support.
\newpage

\appendix

\section{}
The object of this appendix is to obtain the spectral decomposition of the operators $~ \frac{\partial^2}{\partial x^2} \kappa(x) $, and $~ \frac{\partial ~}{\partial x} \kappa(x) \frac{\partial ~}{\partial x}$,  where  the thermal diffusivity is $~ \kappa(x)  ~ =~ \kappa_1 ~ \theta(-x) ~+ ~ \kappa_2 ~ \theta(x) ~\theta(L-x) ~ + ~ \kappa_1 ~ \theta(x-L)$, with the condition that the eigenfunctions are a superposition of plane waves in the limit $~ \left|x \right| ~\rightarrow ~ \infty $. This corresponds to the setup depicted in Fig. 1 with semi-infinite reservoirs.\\
\indent Consider  the eigenvalue equation
\begin{equation}\label{A1}
\frac{\partial^2}{\partial x^2} \kappa(x)~ P_k(x) ~ =~ -  k^2~ P_k(x).
\end{equation}
We require that $~ \kappa(x)~ P_k(x) ~$ and its first derivative  be continuous at $~ x=0,L $, and that the  $~ P_k(x) ~$ form an orthonormal set (in the Dirac sense, that is  we require that $~ \int_{-\infty}^{\infty} dx ~ \kappa(x)~ P_k(x) ~ P_{k^{'}}(x) ~ = ~ \delta(k-k^{'}) ~$). The eigenfunctions can be written as
\begin{eqnarray}\label{A2}
P_k(x) & = & \frac{1}{\kappa_1} \left[ a_k ~ \cos \frac{kx}{\sqrt{\kappa_1}} ~ +~ b_k ~\sqrt{\kappa_1}~ \sin \frac{kx}{\sqrt{\kappa_1}}  \right] \theta(-x) ~  \nonumber \\
& + & \frac{1}{\kappa_2} \left[ c_k ~ \cos \frac{kx}{\sqrt{\kappa_2}} ~ +~ d_k ~\sqrt{\kappa_2}~ \sin \frac{kx}{\sqrt{\kappa_2}}  \right] ~ \theta(x) ~ \theta(L-x) ~  \nonumber \\
& + & \frac{1}{\kappa_1} \left[ e_k ~ \cos \frac{k(x-L)}{\sqrt{\kappa_1}} ~ +~ f_k ~\sqrt{\kappa_1}~ \sin \frac{k(x-L)}{\sqrt{\kappa_1}}  \right] ~ \theta(x-L). 
\end{eqnarray}
First, we implement the continuity conditions at $~ x=0 $, and obtain the equalities
\begin{eqnarray}\label{A3}
c_k & = & a_k, \nonumber \\
b_k & = & d_k.
\end{eqnarray}
The continuity conditions at $~ x=L$, yield the result
\begin{equation}\label{A4}
\left(\begin{array}{c} e_k\\ f_k \end{array} \right) ~ = ~ \left(\begin{array}{cc} \cos \frac{kL}{\sqrt{\kappa_2}} & \sqrt{\kappa_2}~\sin \frac{kL}{\sqrt{\kappa_2}}  \\  -\frac{1}{\sqrt{\kappa_2}} ~\sin \frac{kL}{\sqrt{\kappa_2}} &\cos \frac{kL}{\sqrt{\kappa_2}} \end{array}  \right)~  \left(\begin{array}{c} c_k\\ d_k \end{array} \right).
\end{equation}
Finally, the orthonormalization condition is
\begin{eqnarray}\label{A5}
&& \int_{-\infty}^{\infty} dx ~ \kappa(x)~ P_k(x) ~ P_{k^{'}}(x) ~ = ~ \frac{\pi}{2 \sqrt{\kappa_1}} \left\{ (a_k^2 ~+~ \kappa_1 b_k^2 ~+~ e_k^2 ~+~ \kappa_1 f_k^2)  ~ \delta(k-k^{'}) ~  \right. \nonumber \\
&& \left.\ \ \ \ \  \ \ \ +  ~ (a_k a_{-k} ~-~ \kappa_1 b_k b_{-k} ~+~ e_k e_{-k} ~-~ \kappa_1 f_k f_{-k})  ~ \delta(k+k^{'})  \right\} ~= ~ \delta(k-k^{'})
\end{eqnarray}
Combining Eqs. (A3-A5), we obtain
\begin{equation}\label{A6}
b_k ~ = ~ \alpha_{\left|k \right|} ~ a_k, 
\end{equation}
with
\begin{equation}\label{A7}
\alpha_{\left|k \right|}^2 ~ = ~ \frac{1}{\kappa_1} ~ \frac{1 ~+~\cos^2 \frac{kL}{\sqrt{\kappa_2}} ~+~  \frac{\kappa_1}{\kappa_2} ~ \sin^2 \frac{kL}{\sqrt{\kappa_2}}}{1 ~+~\cos^2 \frac{kL}{\sqrt{\kappa_2}} ~+~  \frac{\kappa_2}{\kappa_1} ~ \sin^2 \frac{kL}{\sqrt{\kappa_2}}}.
\end{equation}
We are mainly interested in the limit $~  \epsilon ~= ~ \sqrt{\frac{\kappa_2}{\kappa_1}} ~\rightarrow~ 0 ~$; then 
\begin{equation}\label{A8}
\alpha_{\left|k \right|}^2 ~ {\sim} ~ \frac{1}{\kappa_1} ~ + ~  \frac{1}{\kappa_2} ~ \frac{\sin^2 \frac{kL}{\sqrt{\kappa_2}}}{ 1 ~ + ~ \cos^2 \frac{kL}{\sqrt{\kappa_2}}},
\end{equation}
when $~ \epsilon \rightarrow 0.$ Note that if $~ \frac{kL}{\sqrt{\kappa_2}}~ \neq ~ n~\pi ~$ for  any $ \ n~ \in ~{\cal{Z}} \ $ then all coefficients $~ a_k,...,f_k ~$ are at most of order $~\epsilon$. On the other hand, if  $~ \frac{kL}{\sqrt{\kappa_2}}~ =~ n \pi ~+~ \eta $, with  $~\eta \ll 1~$ for some $ \ n~ \in ~{\cal{Z}} \ $, then  all coefficients are at most of order $~\epsilon~$ except $~ a_k$, which has a singular behavior,  
\begin{equation}\label{A9}
a_k^2 ~ = ~  \frac{\sqrt{\kappa_2}}{\pi}~ \frac{\epsilon}{(2 \epsilon^2 + \eta^2) ~ - ~ \frac{1}{\sqrt{2}}~ \eta ~(2 \epsilon^2 + \eta)^{\frac{1}{2}}}.
\end{equation}
For $~ \epsilon \rightarrow 0 ~$ Eq. (A9) is an asymmetric representation of a delta function. That is,
\begin{equation}\label{A10}
a_k^2  ~ \sim ~ \sqrt{\kappa_2} ~ \delta(\eta), 
\end{equation}
when $~\epsilon \rightarrow 0.$ \\
\indent Thus, to lowest order in $~ \epsilon $, the eigenfunctions of the operator $~ \frac{\partial^2}{\partial x^2} \kappa(x) ~$ are 
\begin{equation}\label{A11}
P_k(x) ~ = ~ \frac{a_k}{\kappa_2} ~ \cos \frac{kx}{\sqrt{\kappa_2}} ~\theta(x) ~ \theta(L-x),
\end{equation}
with
\begin{equation}\label{A12}
a_k^2 ~\approx ~ \sqrt{\kappa_2} ~ \sum_{n= -\infty}^{\infty} ~ \delta(\frac{kL}{\sqrt{\kappa_2}} ~-~ n \pi) ~= ~ \frac{\kappa_2}{L} ~  \sum_{n= -\infty}^{\infty} ~ \delta(k ~-~ n ~ \frac{\pi \sqrt{\kappa_2}}{L}).
\end{equation}
\indent For the second operator, the eigenvalue equation is
\begin{equation}\label{A13}
\frac{\partial ~}{\partial x} \kappa(x) \frac{\partial ~}{\partial x} ~ Q_k(x) ~ = ~ - ~ k^2 ~ Q_k(x),
\end{equation}
with the conditions that $~Q_k(x)~$ and $~\kappa(x) \frac{\partial ~}{\partial x} ~ Q_k(x)~$ are continuous at $~x =0,L$. We further require that the set of eigenfunctions be orthonormal in the Dirac sense $~ \int_{-\infty}^{\infty}~dx~ Q_k(x) Q_{k^{'}}(x) ~ = ~ \delta(k-k^{'})$.\\
The solution of Eq. (A13) can be written as
\begin{eqnarray}\label{A14}
Q_k(x) & = &  \left[ a_k ~ \cos \frac{kx}{\sqrt{\kappa_1}} ~ +~ \frac{b_k}{\sqrt{\kappa_1}}~ \sin \frac{kx}{\sqrt{\kappa_1}}  \right] \theta(-x) ~  \nonumber \\
& + &  \left[ c_k ~ \cos \frac{kx}{\sqrt{\kappa_2}} ~ +~ \frac{d_k}{\sqrt{\kappa_2}}~ \sin \frac{kx}{\sqrt{\kappa_2}}  \right] ~ \theta(x) ~ \theta(L-x) ~  \nonumber \\
& + &  \left[ e_k ~ \cos \frac{k(x-L)}{\sqrt{\kappa_1}} ~ +~ \frac{f_k}{\sqrt{\kappa_1}}~ \sin \frac{k(x-L)}{\sqrt{\kappa_1}}  \right] ~ \theta(x-L). 
\end{eqnarray}
Using the continuity and orthonormalization conditions, we have
\begin{eqnarray}\label{A15}
c_k & = & a_k \nonumber \\
b_k & = & d_k \nonumber \\ \nonumber \\
\left(\begin{array}{c} e_k\\ f_k \end{array} \right) & = & \left(\begin{array}{cc} \cos \frac{kL}{\sqrt{\kappa_2}} & \frac{1}{\sqrt{\kappa_2}}~\sin \frac{kL}{\sqrt{\kappa_2}}  \\  -\sqrt{\kappa_2} ~\sin \frac{kL}{\sqrt{\kappa_2}} &\cos \frac{kL}{\sqrt{\kappa_2}} \end{array}  \right)~  \left(\begin{array}{c} c_k \\ d_k \end{array} \right) \nonumber \\ \nonumber \\
a_k & = &  \alpha_{\left|k\right|} ~b_k \nonumber \\
\alpha_{\left|k\right|}^2 & = & \frac{1}{\kappa_1} ~ \frac{1 ~+~\cos^2 \frac{kL}{\sqrt{\kappa_2}} ~+~  \frac{\kappa_1}{\kappa_2} ~ \sin^2 \frac{kL}{\sqrt{\kappa_2}}}{1 ~+~\cos^2 \frac{kL}{\sqrt{\kappa_2}} ~+~  \frac{\kappa_2}{\kappa_1} ~ \sin^2 \frac{kL}{\sqrt{\kappa_2}}}.
\end{eqnarray}
In the limit $~  \epsilon ~= ~ \sqrt{\frac{\kappa_2}{\kappa_1}} ~\rightarrow~ 0 ~$ we have
\begin{equation}\label{A16}
Q_k(x) ~ \approx ~ \frac{b_k}{\kappa_2} ~ \sin \frac{kx}{\sqrt{\kappa_2}} ~ \theta(x) \theta(L-x),
\end{equation}
with 
\begin{equation}\label{A17}
b_k^2 ~ \approx~ \frac{\kappa_2}{L} ~  \sum_{n= -\infty}^{\infty} ~ \delta(k ~-~ n ~ \frac{\pi \sqrt{\kappa_2}}{L}), 
\end{equation}
when $~ \epsilon ~ \rightarrow ~0.$

\section{}
Here we compute the values of  $~ \nu_{\{s\}}~$ and of $~ \sum_{\{\sigma\}} \sigma_i ~ \left<A\right>_{\{s\} \rightarrow \{\sigma\}} ~ $ for the automaton described in Section III.\\
\indent We first evaluate the number of elements in the equivalence class of configurations with $~\sum_k s_k ~$ particles:
\begin{eqnarray*}
&& \hspace{-0.7cm}\nu_{\{s\}} ~ = ~ \sum_{\{\sigma\}} ~ \delta(\sum_k s_k, \sum_k \sigma_k)  ~= \nonumber \\
&& \hspace{-0.6cm} = ~  \sum_{\{\sigma\}} \frac{1}{2\pi} \int_0^{2\pi} dx ~ \exp\left\{ i \sum_{k=0}^3 (s_k-\sigma_k) ~ x \right\} ~ = ~ \frac{1}{2\pi}~ \int_0^{2\pi} dx ~ \sum_{\{\sigma\}} ~\prod_{k=0}^3 \exp\left\{ i~(s_k-\sigma_k)~x \right\} ~  \nonumber \\
&& \hspace{-0.6cm} = ~ \frac{1}{2\pi}~ \int_0^{2\pi} dx ~  \prod_{k=0}^3  \left[e^{i ~s_k x}~\sum_{\sigma_k=0}^1 e^{-i~ \sigma_k x }\right] ~ =  ~ \frac{1}{2\pi}~ \int_0^{2\pi} dx ~ \exp\left\{i~ \sum_k s_k x \right\} ~  \left( 1 + e^{-ix} \right)^4 ~   \nonumber \\
&& \hspace{-0.6cm} = ~  \sum_{l=0}^4 ~ \left( \begin{array}{c} 4 \\ l \end{array} \right) ~  \frac{1}{2 \pi}~ \int_0^{2\pi} dx ~  \exp\left\{i ~(\sum_k s_k -l) x \right\} ~ = ~ \sum_{l=0}^4 \left(\begin{array}{c} 4 \\ l \end{array} \right) ~ \delta(l, \sum_ks_k) ~ = ~   \left(\begin{array}{c} 4 \\ \sum_k s_k \end{array} \right). \\
\end{eqnarray*}
Thus, the number of  configurations with a given number of particles is
\begin{equation}\label{B1}
\nu_{\{s\}} ~= ~ \left(\begin{array}{c} 4 \\ \sum_k s_k \end{array} \right),
\end{equation}
i.e., the number different ways of placing $~ \sum_k s_k ~$ particles in the four different channels. 
\indent We now evaluate the numerator of the second expression, which is the number of configurations with channel $i$ occupied
\begin{eqnarray}\label{B2}
&&\hspace{-0.7cm} \sum_{\{\sigma\}} ~\sigma_i~ \delta(\sum_k s_k, \sum_k \sigma_k) ~ = ~ \frac{1}{2\pi}~ \int_0^{2\pi} dx~ \exp\left\{i ~\sum_k s_k ~x \right\} ~\left(1+e^{-i x} \right)^3 ~  \prod_{k \neq i}  \sum_{\sigma_i=0}^{1} ~ \sigma_i ~ e^{-i \sigma_i x} ~  ~ \nonumber \\
&&   \ \ \ \ \ \ \ \ =  ~ \frac{1}{2\pi}~ \int_0^{2\pi} dx~ \exp\left\{i ~\sum_k s_k ~ x \right\}  \left(1+e^{-i x} \right)^3 ~ e^{-i x} ~= ~  \left(\begin{array}{c} 3 \\ \sum_k s_k -1 \end{array} \right). \nonumber \\
\end{eqnarray}
This result could have been obtained straightforwardly by arguing that the quantity calculated is equal to the number of different ways of placing the remaining $~ \sum_k s_k -1 ~$ particles in the $~3~$ available channels, given that  channel $i$ is already occupied by one particle.\\
\indent Combining  (B1) and (B2), we have
\begin{equation}\label{B3}
\sum_{\{\sigma\}} \sigma_i ~ A_{\{s\} \rightarrow \{\sigma\}}  ~ = ~ \frac{\sum_{\{\sigma\}} \sigma_i~ \delta(\sum_k s_k, \sum_k \sigma_k)}{\sum_{\{\sigma\}} ~ \delta(\sum_k s_k, \sum_k \sigma_k)} ~ = ~\frac{\sum_ks_k}{4},
\end{equation}
which just states the fact that after a collision, the average occupation of a channel is proportional to the number of particles participating in the collision.

\section{}
In a closed system at equilibrium  long-range correlations appear rather trivially as  a consequence of global conservation laws. For the automata under consideration, it has been verified (see Fig. 5(b)) that  equilibrium correlations of the density fluctuations  exhibit  a long-range   contribution, which is constant, and whose integrated value is equal in magnitude and opposite in sign to the local equilibrium term, so  that the total number of particles  be conserved.   \\
\indent Consider an automaton whose lattice $~{\cal{L}}~$  contains  $V$ nodes. There are $b$ equivalent channels per node. The particle density (number of particles per node) is $\rho$. We consider periodic boundary conditions, so that the system is closed (i. e. the total number of particles $~ N = \rho V~$ is conserved). Let $ ~\{ n(\cdot) \} ~= ~\{n_i({\bf r});~ {\bf r} \in {\cal{L}},~ i = 0,1... (b-1) \} $ denote the automaton configuration. The equilibrium  probability of having the configuration $\{n(\cdot) \} $ is
\begin{equation}\label{C1}
P\left[ \{n(\cdot) \} \right] ~ = ~ \prod_{{\bf r}} ~ \prod_{i} ~ p\left(n_i({\bf r}) \right) ~ \delta\left(\sum_{{\bf r}} \sum_i ~ n_i({\bf r}) ~ - ~ N \right),
\end{equation}
with $~ p\left(n_i({\bf r}) = 1 \right) ~ = ~ \theta $, $~ p\left(n_i({\bf r}) = 0 \right) ~ = ~ 1- \theta $. The parameter $\theta$ is determined by normalization:
\begin{eqnarray}\label{C2}
\sum_{\{n(\cdot)\}} ~ P\left[ \{n(\cdot) \} \right] & = & \sum_{\{n(\cdot)\}}
  \prod_{{\bf r}} ~ \prod_{i} ~ p\left(n_i({\bf r}) \right) ~ \frac{1}{2\pi} \int_0^{2 \pi} ~ dx ~ \exp\left\{ i ~x \left( \sum_{{\bf r}} \sum_i n_i({\bf r}) \right) ~-~ N\right\} ~  \nonumber \\
& = &  \frac{1}{2\pi} ~ \int_0^{2 \pi} ~ dx ~ e^{- i x N} ~  \prod_{{\bf r}} ~ \prod_{i} ~ \sum_{n_i({\bf r})=0}^1 ~p(n_i({\bf r})) e^{ixn_i({\bf r})} ~  \nonumber \\
& = &  \frac{1}{2\pi} ~ \int_0^{2 \pi} ~ dx ~ e^{ -i x N} ~ \prod_{{\bf r}} ~ \prod_{i} ~ \left[ \theta e^{ix} + (1-\theta) \right] ~  \nonumber \\
& = &   \frac{1}{2\pi} ~ \int_0^{2 \pi} ~ dx ~ e^{ -i x N} ~  \left[ \theta e^{ix} + (1-\theta) \right]^{Vb} ~  ~ \nonumber \\
& = & \left( \begin{array}{c} Vb \\ N \end{array} \right) ~ \theta^N ~ (1-\theta)^{Vb-N} ~ = 1.
\end{eqnarray}
The last equality yields  $~ \theta ~$ as an implicit  function of the automaton parameters $~V,b~$ and of the occupation $~N$. \\
\indent In a similar way, the average occupation per channel is
\begin{equation}\label{C3}
\left<n_i({\bf r}) \right> ~ = ~ \sum_{\{n(\cdot)\}} ~ n_i({\bf r}) ~ P\left[ \{n(\cdot) \} \right] ~ = ~  \frac{1}{2\pi} ~ \int_0^{2 \pi}  dx ~ e^{ -i x N} ~  \left[ \theta e^{ix} + (1-\theta) \right]^{Vb -1} ~ \theta ~e^{ix} ~ = ~ ~\frac{\rho}{b} 
\end{equation}
and the two particle distributions, for $~ ({\bf r},i) ~ \neq ({\bf r^{'}},j) $, are
\begin{eqnarray}\label{C4}
\left<n_i({\bf r}) n_j({\bf r^{'}})\right> & = & \sum_{\{n(\cdot)\}} ~ n_i({\bf r}) ~ n_j({\bf r^{'}}) ~  P\left[ \{n(\cdot) \} \right]  ~ \nonumber \\
& = &  \frac{1}{2\pi}  \int_0^{2 \pi}  dx ~ e^{ -i x N} ~  \left[ \theta e^{ix} + (1-\theta) \right]^{Vb -2} ~ \theta^2 ~e^{2ix} ~ = ~ ~\frac{\rho}{b} ~ \frac{N-1}{Vb-1}. 
\end{eqnarray}
\indent By contrast, in an automaton which is maintained at equilibrium by contact with particle reservoirs, the total number of particles is conserved only on average, and the  equilibrium distribution is 
\begin{equation}\label{C5}
P\left[ \{n(\cdot) \} \right] ~ = ~ \prod_{{\bf r}} ~ \prod_{i} ~ p\left(n_i({\bf r}) \right) ~ \delta\left(\sum_{{\bf r}} \sum_i ~ n_i({\bf r}) ~ - ~ N \right),
\end{equation}
with $~ p\left(n_i({\bf r} = 1) \right) ~ = ~ \frac{\rho}{b} $, $~ p\left(n_i({\bf r} = 0) \right) ~ = ~ 1- \frac{\rho}{b} $, which guarantees normalization. The average occupation per channel is also $ ~ \left< n_i({\bf r}) \right> ~ = ~ \frac{\rho}{b} $, but  correlations differ from the previous case (compare Figs. 5(a)  and 5(b)). For  $~ ({\bf r},i) ~ \neq ({\bf r^{'}},j) $, the two particle distribution factorizes  $~ \left<n_i({\bf r}) n_j({\bf r^{'}})\right> ~ = ~\left<n_i({\bf r}) \right> \left< n_j({\bf r^{'}} )\right> ~ = ~ (\frac{\rho}{b})^2 $, which implies that the correlations are strictly short-ranged, in a system in contact with particle reservoirs.

\bibliographystyle{unsrt}

\begin{thebibliography}{10}

\bibitem{law89}
B.~M. Law and J.~V. Sengers,
\newblock {\em J. Stat. Phys.} {\bf 57}, 531 (1989).

\bibitem{law90}
B.~M. Law, P.~N. Segr\`{e}, R.~W. Gammon and J.~V. Sengers,
\newblock {\em Phys. Rev. A} {\bf 41}, 816 (1990).

\bibitem{segre92}
P.~N. Segr\`{e}, R.~W. Gammon, J.~V. Sengers and B.~M. Law,
\newblock {\em Phys. Rev. A} {\bf 45}, 714 (1992).

\bibitem{li94}
W.~B. Li, P.~N. Segr\`{e}, R.~W. Gammon and J.~V. Sengers,
\newblock {\em Physica A} {\bf 204}, 399 (1994).

\bibitem{kirkpatrick82}
T.~R. Kirkpatrick, E.~G.~D. Cohen and J.~R. Dorfman,
\newblock {\em Phys. Rev. A} {\bf 26}, 950, 972, 995 (1982).

\bibitem{ronis80}
D.~Ronis, I.~Procaccia and J.~Machta,
\newblock {\em Phys. Rev. A} {\bf 22}, 714 (1980), and references therein.

\bibitem{ronis82}
D.~Ronis and I.~Procaccia,
\newblock {\em Phys. Rev. A} {\bf 26}, 1812 (1982).

\bibitem{fox82}
R.~F. Fox,
\newblock {\em J. Phys. Chem.} {\bf 86} 2812 (1982).

\bibitem{tremblay84}
A.-M.~S. Tremblay, in
\newblock {\em Recent Developments in Nonequilibrium Thermodynamics}, J.
  Casas-V\'{a}zquez, D. Jou and G. Lebon eds. (Springer-Verlag, Berlin, 1984).

\bibitem{perezmadrid86}
A.~P\'erez Madrid and J.~M. Rub\'{\i},
\newblock {\em Phys. Rev. A} {\bf 33}, 2716 (1986).

\bibitem{hwa89}
T.~Hwa and M.~Kardar,
\newblock {\em Phys. Rev. Lett.} {\bf 62}, 1813 (1989).

\bibitem{garrido90}
P.~L. Garrido, J.~L. Lebowitz, C.~Maes and H.~Spohn,
\newblock {\em Phys. Rev. A} {\bf 42}, 1954 (1990).

\bibitem{beijeren90}
H.~van Beijeren,
\newblock {\em J. Stat. Phys.} {\bf 60}, 845 (1990).

\bibitem{grinstein91}
G.~Grinstein,
\newblock {\em J. Appl. Phys.} {\bf 69}, 5441 (1991).

\bibitem{pagonabarraga94}
I.~Pagonabarraga and J.~M. Rub\'{\i},
\newblock {\em Phys. Rev. E} {\bf 49}, 267 (1994).

\bibitem{dorfman94}
J.~R. Dorfman, T.~R. Kirkpatrick  and J.~V. Sengers,
\newblock {\em Ann. Rev. Phys. Chem.} {\bf 45}, 213 (1994).

\bibitem{lebowitz88}
J.L. Lebowitz, E.~Presutti and H.~Spohn,
\newblock {\em J. Stat. Phys.} {\bf 51}, 841 (1988).

\bibitem{grosfils92}
P.~Grosfils, J.~P. Boon and P.~Lallemand,
\newblock {\em Phys. Rev. Lett.} {\bf 68}, 1077 (1992).

\bibitem{bussemaker95}
H.~J. Bussemaker, M.~H. Ernst and J.~W. Dufty.
\newblock {\em J. Stat. Phys.} {\bf 78}, 1521 (1995).

\bibitem{kawasaki66}
K.~Kawasaki,
\newblock {\em Phys. Rev.} {\bf 145}, 225 (1966).

\bibitem{spohn83}
H.~Spohn,
\newblock {\em J. Phys. A: Math. Gen.} {\bf 16}, 4275 (1983).

\bibitem{nicolis84}
G.~Nicolis and M.~Malek Mansour,
\newblock {\em Phys. Rev. A} {\bf 29} 2845 (1984).

\bibitem{garcia87}
A.~L. Garcia, M.~Malek Mansour, G.~C. Lie and E.~Clementi,
\newblock {\em J. Stat. Phys.} {\bf 47}, 209 (1987).

\bibitem{dubrulle90}
B.~Dubrulle, U.~Frisch, M.~H\'enon and J.~P. Rivet,
\newblock {\em J. Stat. Phys.} {\bf 59}, 1187 (1990).

\bibitem{ernst95}
M.~H. Ernst and H.~J. Bussemaker,
\newblock {\em  J. Stat. Phys.} {\bf 81}, 515 (1995).

\bibitem{procaccia79}
I.~Procaccia, D.~Ronis, M.~A. Collins, J.~Ross and I.~Oppenheim,
\newblock {\em Phys. Rev. A} {\bf 19}, 1290 (1979).

\bibitem{foot0}
It is convenient to use $~ L_x+1~$ nodes because fixed boundary conditions will
  be imposed at $x=0$ and $x=L_x$. Along the $Y$ direction we impose periodic
  boundary conditions, identifying nodes $y=0$ with nodes $y=L_y$.

\bibitem{boon95}
J.~P. Boon, D.~Dab, R.~Kapral and A.~Lawniczak, Lattice gas automata for reactive systems, \newblock to appear in {\em  Phys. Rep.}

\bibitem{foot1}
$\pm 0.002$ is an estimate of the dispersion of $D$ for different values of
  $k$.

\bibitem{landau}
L.~D. Landau and E.~M. Lifshitz, 
\newblock {\em Fluid Mechanics}, Chapter XVII (Pergamon Press, London, 1959).

\bibitem{ernst90}
M.~H. Ernst, in
\newblock {\em Liquids, freezing and the glass transition}, D. Levesque, J. P.
  Hansen and J. Zinn-Justin eds (Elsevier, Amsterdam, 1990).

\bibitem{macquarrie}
D.~A. McQuarrie,
\newblock {\em Statistical Mechanics} (Harper and Row, New York, 1973).

\end{thebibliography}

\end{document}